\documentstyle[12pt]{article}
\begin{document}

\def\ds{\displaystyle}
\def\beq{\begin{equation}}
\def\eeq{\end{equation}}
\def\bea{\begin{eqnarray}}
\def\eea{\end{eqnarray}}
\def\beeq{\begin{eqnarray}}
\def\eeeq{\end{eqnarray}}
\def\ve{\vert}
\def\vel{\left|}
\def\ver{\right|}
\def\nnb{\nonumber}
\def\ga{\left(}
\def\dr{\right)}
\def\aga{\left\{}
\def\adr{\right\}}
\def\lla{\left<}
\def\rra{\right>}
\def\rar{\rightarrow}
\def\nnb{\nonumber}
\def\la{\langle}
\def\ra{\rangle}
\def\ba{\begin{array}}
\def\ea{\end{array}}
\def\tr{\mbox{Tr}}
\def\ssp{{\Sigma^{*+}}}
\def\sso{{\Sigma^{*0}}}
\def\ssm{{\Sigma^{*-}}}
\def\xis0{{\Xi^{*0}}}
\def\xism{{\Xi^{*-}}}
\def\qs{\la \bar s s \ra}
\def\qu{\la \bar u u \ra}
\def\qd{\la \bar d d \ra}
\def\qq{\la \bar q q \ra}
\def\gGgG{\la g^2 G^2 \ra}
\def\q{\gamma_5 \not\!q}
\def\x{\gamma_5 \not\!x}
\def\g5{\gamma_5}
\def\sb{S_Q^{cf}}
\def\sd{S_d^{be}}
\def\su{S_u^{ad}}
\def\ss{S_s^{??}}
\def\ll{\Lambda}
\def\lb{\Lambda_b}
\def\sbp{{S}_Q^{'cf}}
\def\sdp{{S}_d^{'be}}
\def\sup{{S}_u^{'ad}}
\def\ssp{{S}_s^{'??}}
\def\sig{\sigma_{\mu \nu} \gamma_5 p^\mu q^\nu}
\def\fo{f_0(\frac{s_0}{M^2})}
\def\ffi{f_1(\frac{s_0}{M^2})}
\def\fii{f_2(\frac{s_0}{M^2})}
\def\O{{\cal O}}
\def\sl{{\Sigma^0 \Lambda}}
\def\es{\!\!\! &=& \!\!\!}
\def\ar{&+& \!\!\!}
\def\ek{&-& \!\!\!}
\def\cp{&\times& \!\!\!}
\def\se{\!\!\! &\simeq& \!\!\!}
\def\hml{\hat{m}_{\ell}}
\def\rr{\hat{r}_{\Lambda}}
\def\ss{\hat{s}}


\renewcommand{\textfraction}{0.2}    
\renewcommand{\topfraction}{0.8}

\renewcommand{\bottomfraction}{0.4}
\renewcommand{\floatpagefraction}{0.8}
\newcommand\mysection{\setcounter{equation}{0}\section}

\def\baeq{\begin{appeq}}     \def\eaeq{\end{appeq}}
\def\baeeq{\begin{appeeq}}   \def\eaeeq{\end{appeeq}}
\newenvironment{appeq}{\beq}{\eeq}
\newenvironment{appeeq}{\beeq}{\eeeq}
\def\bAPP#1#2{
 \markright{APPENDIX #1}
 \addcontentsline{toc}{section}{Appendix #1: #2}
 \medskip
 \medskip
 \begin{center}      {\bf\LARGE Appendix #1 :}{\quad\Large\bf #2}
\end{center}
 \renewcommand{\thesection}{#1.\arabic{section}}
\setcounter{equation}{0}
        \renewcommand{\thehran}{#1.\arabic{hran}}
\renewenvironment{appeq}
  {  \renewcommand{\theequation}{#1.\arabic{equation}}
     \beq
  }{\eeq}
\renewenvironment{appeeq}
  {  \renewcommand{\theequation}{#1.\arabic{equation}}
     \beeq
  }{\eeeq}
\nopagebreak \noindent}

\def\eAPP{\renewcommand{\thehran}{\thesection.\arabic{hran}}}

\renewcommand{\theequation}{\arabic{equation}}
\newcounter{hran}
\renewcommand{\thehran}{\thesection.\arabic{hran}}

\def\bmini{\setcounter{hran}{\value{equation}}
\refstepcounter{hran}\setcounter{equation}{0}
\renewcommand{\theequation}{\thehran\alph{equation}}\begin{eqnarray}}
\def\bminiG#1{\setcounter{hran}{\value{equation}}
\refstepcounter{hran}\setcounter{equation}{-1}
\renewcommand{\theequation}{\thehran\alph{equation}}
\refstepcounter{equation}\label{#1}\begin{eqnarray}}


\newskip\humongous \humongous=0pt plus 1000pt minus 1000pt
\def\caja{\mathsurround=0pt}


\title{
 {\small \begin{flushright}
IPM/P-2007/049\\
\today
\end{flushright}}
       {\Large
                 {\bf
 Double Lepton Polarization
in $\Lambda_b \rar \Lambda \ell^+ \ell^-$ Decay in the Standard
Model with Fourth Generations Scenario
                 }
         }
      }

\author{\vspace{1cm}\\
{\small F. Zolfagharpour$^1$\thanks {e-mail:
zolfagharpour@uma.ac.ir}\,\,, V.Bashiry$^2$\thanks {e-mail:
bashiry@ipm.ir}\,\,,
} \\
{\small $^1$ Department of physics, The university of Mohaghegh
Ardabili , P.O. Box 179, Ardabil, Iran}\\ {\small $^2$ Institute
for Studies in Theoretical Physics and Mathematics (IPM),}\\
{\small P.O. Box 19395-5531, Tehran, Iran }\\}
\date{}
\begin{titlepage}
\maketitle
\thispagestyle{empty}

\begin{abstract}
This study investigates the influence of the fourth generation
quarks on the double lepton polarizations in $\Lambda_b \rar \Lambda
\ell^+ \ell^-$ decay by taking $|V^\ast_{t's}V_{t'b}|=
0.005,0.01,0.02,0.03 $ with phase $\{60^\circ,90^\circ,120^\circ\}$.
We will try to obtain a constrain on the mass of the 4th generation
top like quark $t'$, which is consistent with the $b\to
s\ell^+\ell^-$ rate . With the above mentioned parameters, we will
try to show that the double lepton($\mu, \, \tau$) polarizations are
quite sensitive to the existence of fourth generation. It can serve
as a good tool to search for new physics effects, precisely, to
search for the fourth generation quarks($t',\, b')$ via its indirect
manifestations in loop diagrams.
\end{abstract}

~~~PACS numbers: 12.60.--i, 13.30.--a, 14.20.Mr
\end{titlepage}

\section{Introduction}
While the Standard Model (SM) provides a very good description of
phenomena observed by experiments, it is still an incomplete theory.
The problem is that the Standard Model can't explain why some
particles exist as they do. Another question concerns the fact that
there are 3 pairs of quarks and 3 pairs of leptons. Each "set" of
these particles is called a generation (a.k.a. family). Therefore,
the up/down quarks are first generation quarks, while the electron
and e--neutrino are first generation leptons. In the every-day world
we observe only the first-generation particles (electrons, e-
neutrinos, and up/down quarks). Why does the natural world, "need"
the two other generations?  Are there  3 generations or more?
Nothing in the standard model itself fixes the number of quarks and
leptons that can exist. Since the first three generations are full,
any new quarks and leptons would be members of a "fourth
generation". In this sense, SM may be treated as an effective theory
of fundamental interactions rather than fundamental particles. The
Democratic Mass Matrix approach \cite{harari}, which is quite
natural in the SM framework, may be considered as the interesting
step in true direction. It is intriguing that Flavors Democracy
favors the existence of the fourth SM family \cite{datta,
celikel,Sultansoy:2000dm}. Any study related to the decay of the 4th
generation quarks or indirect effects of those in FCNC requires the
choice of the quark masses which are not free parameter, rather they
are constrained by the experimental value of
 $\rho$ and $S$ parameters \cite{Sultansoy:2000dm}. The $\rho$
 parameter, in terms of the transverse part of the W--and  Z--boson
 self energies at zero momentum transfer, is given in \cite{Djouadi},
 \bea \rho=\frac{1}{1-\Delta\rho};\,\,\ \Delta\rho=\frac{\Pi_{ZZ}(0)}
 {M_Z^2}-\frac{\Pi_{WW}(0)}{M_W^2},\eea
 the common mass of the fourth generation quarks ($m_{t'}$) lies between
320 GeV and 730 GeV considering the experimental value of
$\rho=1.0002^{+0.0007}_{-0.0004}$ \cite{PDG}. The last value is
close to upper limit on heavy quark mass, $m_q\leq 700$ GeV
$\approx 4m_t$, which follows from partial-wave unitarity at high
energies \cite{chanowitz}. It should be noted that with preferable
value $a\approx g_w$, Flavor Democracy predicts $m_{t'}\approx 8
m_w\approx 640$ GeV. The above mentioned values for mass of
$m_{t'}$ disfavors the fifth SM family both because in general we
expect that $m_t\leq m_{t'}\leq m_t^{''}$ and the experimental
values of the $\rho$ and $S$ parameters \cite{Sultansoy:2000dm}
restrict the quark mass up to $700$ GeV.

The study of production, decay channels and LHC signals of the 4th
generation quarks have been continuing. But, one  of the efficient
ways to establish the existence of 4th generation
 is via their indirect manifestations in loop diagrams. Rare decays, induced
 by flavor changing neutral
current (FCNC) $b \rar s(d)$ transitions are at the forefront of our
quest to understand flavor and the origins of CPV, offering one of
the best probes for New Physics (NP) beyond the Standard Model (SM).
Several hints for NP have emerged in the past few years.
For example, a large difference is seen in direct CP asymmetries in
$B\to K\pi$ decays~\cite{HFAG},
\begin{eqnarray}
{\cal A}_{K\pi}
 \equiv A_{\rm CP}(B^0\to K^+\pi^-) = -0.093 \pm 0.015, &&\nonumber\\
{\cal A}_{K\pi^0}
 \equiv A_{\rm CP}(B^+\to K^+\pi^0) = +0.047 \pm 0.026, &&
 \label{data}
\end{eqnarray}
or $\Delta{\cal A}_{K\pi} \equiv {\cal A}_{K\pi^0}-{\cal A}_{K\pi} =
(14\pm 3)\%$~\cite{Barlow}. As this percentage was not predicted
when first measured in 2004, it has stimulated discussion on the
potential mechanisms that it may have been missed in the SM
calculations \cite{BBNS,KLS,BPS05}.

Better known is the mixing-induced CP asymmetry ${\cal S}_f$
measured in a multitude of CP eigenstates $f$. For
penguin-dominated $b \to sq\bar q$ modes, within SM, ${\cal
S}_{sq\bar q}$ should be close to that extracted from $b\to c\bar
cs$ modes. The latter is now measured rather precisely, ${\cal
S}_{c\bar cs}=\sin2\phi_1 = 0.674 \pm 0.026$~\cite{Hazumi}, where
$\phi_1$ is the weak phase in $V_{td}$. However, for the past few
years, data seem to indicate at 2.6$\,\sigma$ significance,
\begin{eqnarray}
\Delta {\cal S} \equiv {\cal S}_{sq\bar q}-{\cal S}_{c\bar cs}\leq
0,
 \label{DelS}
\end{eqnarray}
which has stimulated even more discussions.

The $b \rar s(d) \ell^+ \ell^-$ decays have received considerable
attention as a potential testing ground for the effective
Hamiltonian describing FCNC in B and $\Lambda_b$ decay. This
Hamiltonian contains the one--loop effects of the electroweak
interaction, which are sensitive to the quark$^{,}$s contribution in
the loop \cite{Willey}--\cite{Buras1995}. In addition, there are
important QCD corrections, which have recently been calculated in
the NNLL\cite{NNLL}. Moreover, $b \rar s(d) \ell^+ \ell^-$ decay is
also very sensitive to the new physics beyond SM. New physics
effects manifest themselves in rare decays in two different ways,
either through new combinations to the new Wilson coefficients or
through the new operator structure in the effective Hamiltonian,
which is absent in the SM. A crucial problem in the new physics
search within flavour physics is the optimal separation of new
physics effects from uncertainties. It is well known that inclusive
decay modes are dominated by partonic contributions;
non--perturbative corrections are in general rather
small\cite{Hurth}. Also ratios of exclusive decay modes such as
asymmetries for $B\rar K(~K^\ast,~\rho,~\gamma)~ \ell^+ \ell^-$
decay \cite{R4621}--\cite{bashirychin} are well studied for new
physics search. Here large parts of the hadronic uncertainties
partially cancel out. In this paper, we investigate the possibility
of searching for new physics in the heavy baryon decays $\Lambda_b
\rar \Lambda \ell^+ \ell^-$ using the SM with four generations of
quarks($b',\, t'$). The fourth quark ($t'$), like $u,c,t$ quarks,
contributes in the $b \rar s(d) $ transition at loop level. Note
that, fourth generation effects have been widely studied in baryonic
and semileptonic B decays \cite{Hou:2006jy}--\cite{Turan:2005pf}.
But, there are few works related to the exclusive decays
$\Lambda_b\rightarrow\Lambda l^{+}l^{-}$\cite{Bashiry2007}.

The main problem for the description of exclusive decays is to
evaluate the form factors, i.e., matrix elements of the effective
Hamiltonian between initial and final hadron states. It is well
known that in order to describe baryonic   $\Lambda_b \rar \Lambda
\ell^+ \ell^-$ decay a number of form factors are needed (see for
example \cite{R4629}). However, when heavy quark effective theory
(HQET) is applied, only two independent form factors appear
\cite{R46210}.

It should be mentioned here that for the exclusive decay $\Lambda_b
\rar \Lambda \ell^+ \ell^-$, decay rate, lepton polarization and
heavy($\Lambda_b$) or light($\Lambda$) baryon polarization(readily
measurable) are studied in the SM, the two Higgs doublet model and
using the general form of the effective Hamiltonian, in
\cite{R4629}, \cite{R46211} and \cite{R46212}--\cite{baryonpol},
respectively.

The sensitivity of the forward--backward asymmetry  to the existence
of fourth generation quarks in the $\Lambda_b \rar \Lambda \ell^+
\ell^-$ decay is investigated in \cite{Turan:2005pf} and it is
obtained that the forward--backward asymmetry is very sensitive to
the fourth generation parameters ($m_{t'}$, $V_{t'b}V^*_{t's}$ ). In
this connection, it is natural to ask whether double lepton
polarizations are sensitive to the fourth generation parameters, in
the "heavy baryon $\rar$ light baryon $\ell^+ \ell^-$" decays. In
the present work, we try to answer to this question.

The paper is organized as follows: In section 2, using the effective
hamiltonian, the general expressions for the matrix element of
$\Lambda_b \rar \Lambda \ell^+ \ell^-$ decay is derived. Section 3
devoted to the calculations of double--lepton polarizations. In
section 4, we investigate the sensitivity of these functions to the
fourth generation parameters ($m_{t'}$, $r_{sb},\,\ \phi_{sb}$ ).

\section{Strategy}

We first consider the standard model contribution. In the SM, the
matrix element of the $\Lambda_b \rar \Lambda \ell^+ \ell^-$ decay
at quark level is described by $b \rar s \ell^+ \ell^-$ transition
for which the effective Hamiltonian at $O(\mu)$ scale can be written
as:
 \bea\label{Hgen} {\cal H}_{eff} &=& \frac{4 G_F}{\sqrt{2}}
V_{tb}V_{ts}^\ast \sum_{i=1}^{10} {\cal C}_i(\mu) \, {\cal
O}_i(\mu)~, \eea where the full set of the operators ${\cal
O}_i(\mu)$ and the corresponding expressions for the Wilson
coefficients ${\cal C}_i(\mu)$ in the SM are given in
\cite{R23}--\cite{R24}. As it has already been noted , the fourth
generation up type quark $t'$ is introduced in the same way as $u,\
c,\ t$ quarks introduce in the SM, and so new operators do not
appear and clearly the full operator set is exactly the same as in
SM. The fourth generation changes the values of the Wilson
coefficients $C_7(\mu),~C_9(\mu)$ and $C_{10}(\mu)$, via virtual
exchange of the fourth generation up type quark $t^\prime$. The
above mentioned Wilson coefficients will explicitly change
  as:
\bea\lambda_t C_i \rightarrow \lambda_t C^{SM}_i+\lambda_{t'}
C^{new}_i~,\eea where $\lambda_f=V_{f b}^\ast V_{fs}$. The unitarity
of the $4\times4$ CKM matrix leads to
\bea\lambda_u+\lambda_c+\lambda_t+\lambda_{t'}=0.\eea\ Since
$\lambda_u=V_{ub}^\ast V_{us}$ is very small in strength compared to
the others . Then $\lambda_t\approx -\lambda_c-\lambda_{t'}$ and
$\lambda_c=V_{c b}^\ast V_{cs}\approx 0.04$ is real by convention.
It follows that \bea \lambda_t C^{SM}_i+\lambda_{t'}
C^{new}_i=\lambda_c C^{SM}_i+\lambda_{t'} (C^{new}_i-C^{SM}_i )\eea
It is clear that, for the $m_{t'}\rar m_t$ or $\lambda_{t'}\rar 0$,
$\lambda_{t'} (C^{new}_i-C^{SM}_i )$ term vanishes, as required by
the GIM mechanism. One can also write $C_i$'s in the following form
\bea\label{c4} C_7^{tot}(\mu) &=& C_7^{SM}(\mu) +
\frac{\lambda_{t'}}
{\lambda_t} C_7^{new} (\mu)~, \nnb \\
C_9^{tot}(\mu) &=& C_9^{SM}(\mu) +  \frac{\lambda_{t'}}
{\lambda_t}C_9^{new} (\mu) ~, \nnb \\
C_{10}^{tot}(\mu) &=& C_{10}^{SM}(\mu) +  \frac{\lambda_{t'}}
{\lambda_t} C_{10}^{new} (\mu)~, \eea
where the last terms in these expressions describe the contributions
of the $t^\prime$ quark to the Wilson coefficients. $\lambda_{t'}$  can be parameterized
as : \bea {\label{parameter}} \lambda_{t'}=V_{t^\prime b}^\ast V_{t^\prime
s}=r_{sb}e^{i\phi_{sb}}\eea

 In deriving Eq. (\ref{c4}) we factored
out the term $V_{tb}^\ast V_{ts}$ in the effective Hamiltonian given
in Eq. (\ref{Hgen}). The explicit forms of the $C_i^{new}$ can
easily be obtained from the corresponding expression of the Wilson
coefficients in SM by substituting $m_t \rar m_{t^\prime}$ (see
\cite{R23,R25}). If the $s$ quark mass is neglected, the above
effective Hamiltonian leads to following matrix element for the $b
\rar s \ell^+ \ell^-$ decay \bea\label{e1} {\cal H}_{eff} &=&
\frac{G\alpha}{2\sqrt{2} \pi}
 V_{tb}V_{ts}^\ast
\Bigg[ C_9^{tot} \, \bar s \gamma_\mu (1-\gamma_5) b \, \bar \ell
\gamma_\mu \ell + C_{10}^{tot} \bar s \gamma_\mu (1-\gamma_5) b \,
\bar \ell \gamma_\mu \gamma_5 \ell \nnb \\
&-& 2  C_7^{tot}\frac{m_b}{q^2} \bar s \sigma_{\mu\nu} q^\nu
(1+\gamma_5) b \, \bar \ell \gamma_\mu \ell \Bigg]~, \eea where
$q^2=(p_1+p_2)^2$ and $p_1$ and $p_2$ are the final leptons
four--momenta. The effective coefficient $C_9^{tot}$ can be written
in the following form \bea C_9^{tot} = C_9 + Y(s')~, \eea where $s'
= q^2 / m_b^2$ and the function $Y(s')$ denotes the perturbative
part coming from one loop matrix elements of four quark operators
and is given\cite{R23,R24}, \bea Y_{per}(s') &=& g(\hat m_c, s') (3
C_1 + C_2 + 3 C_3 + C_4 + 3 C_5 + C_6) \nnb \\&-& \frac{1}{2} g(1,
s') (4 C_3 + 4 C_4 + 3 C_5 + C_6) \nnb \\&-& \frac{1}{2} g(0, s')
(C_3 + 3 C_4) + \frac{2}{9} (3 C_3 + C_4 + 3 C_5 + C_6)~, \eea where
$\hat m_c = \frac{m_c}{m_b} $. The explicit expressions for $g(\hat
m_c, s')$, $g(0, s')$, $g(1, s')$ and the values of $C_i$ in the SM
can be found in Table 1 \cite{R23,R24}.
\begin{table}
\renewcommand{\arraystretch}{1.5}
\addtolength{\arraycolsep}{3pt}
$$
\begin{array}{|c|c|c|c|c|c|c|c|c|}
\hline C_{1} & C_{2} & C_{3} & C_{4} & C_{5} & C_{6} & C_{7}^{SM} &
C_{9}^{SM} & C_{10}^{SM}\\ \hline
-0.248 & 1.107& 0.011& -0.026& 0.007& -0.031& -0.313& 4.344& -4.669\\
\hline
\end{array}
$$
\caption{The numerical values of the Wilson coefficients at $\mu =
m_{b}$ scale within the SM. The corresponding numerical value of
$C_{0}$ is $0.362$.}
\renewcommand{\arraystretch}{1}
\addtolength{\arraycolsep}{-3pt}
\end{table}

In addition to the short distance contribution, $Y_{per}(s')$
receives also long distance contributions, which have their origin
in the real $c\bar c$ intermediate states, i.e., $J/\psi$,
$\psi^\prime$, $\cdots$. The $J/\psi$ family is introduced by the
Breit--Wigner distribution for the resonances through the
replacement \cite{R26}--\cite{R28} \bea Y(s') = Y_{per}(s') +
\frac{3\pi}{\alpha^2} \, C^{(0)} \sum_{V_i=\psi_i} \kappa_i \,
\frac{m_{V_i} \Gamma(V_i \rar \ell^+ \ell^-)} {m_{V_i}^2 - s' m_b^2
- i m_{V_i} \Gamma_{V_i}}~, \eea where $C^{(0)}= 3 C_1 + C_2 + 3 C_3
+ C_4 + 3 C_5 + C_6$. The phenomenological parameters $\kappa_i$ can
be fixed from ${\cal B} (B \rar K^\ast V_i \rar K^\ast \ell^+
\ell^-) = {\cal B} (B \rar K^\ast V_i)\, {\cal B} ( V_i \rar \ell^+
\ell^-)$, where the data for the right hand side is given in
\cite{R29}. For the lowest resonances $J/\psi$ and $\psi^\prime$ one can
use $\kappa = 1.65$ and $\kappa = 2.36$, respectively (see \cite{R30}).

After having an idea of the effective Hamiltonian and the relevant
Wilson coefficients, we now proceed to evaluate the transition
matrix elements for the process $\lb (p_{\lb}) \to \ll (p_\ll)~ l^+
(p_+)l^-(p_-)$. For this purpose, we need to know the marix elements
of the various hadronic currents between initial $\lb$ and the final
$\ll$ baryon, which are parameterized in terms of various form
factors as \bea \langle \ll |\bar s \gamma_\mu b | \lb \rangle & =&
\bar u_\ll \Big[ f_1 \gamma_\mu + i f_2 \sigma_{\mu \nu} q^\nu +f_3
q_\mu \Big]
u_{\lb}\;,\\
\langle \ll |\bar s \gamma_\mu \gamma_5 b | \lb \rangle & =& \bar
u_\ll \Big[ g_1 \gamma_\mu \gamma_5 + i g_2 \sigma_{\mu \nu}\gamma_5
q^\nu + g_3 \gamma_5 q_\mu \Big]
u_{\lb}\;,\\
\langle \ll |\bar s i \sigma_{\mu \nu} q^\nu b | \lb \rangle & =&
\bar u_\ll \Big[ f_1^T \gamma_\mu + i f_2^T \sigma_{\mu \nu} q^\nu
+f_3^T q_\mu \Big]
u_{\lb}\;,\\
\langle \ll |\bar s i \sigma_{\mu \nu}\gamma_5 q^\nu b | \lb \rangle
& =& \bar u_\ll \Big[ g_1^T \gamma_\mu \gamma_5 + i g_2^T
\sigma_{\mu \nu}\gamma_5 q^\nu + g_3^T \gamma_5 q_\mu \Big]
u_{\lb}\;, \eea where $q=p_{\lb}-p_{\ll}=p_++p_-$ is the momentum
transfer, $f_i$ and $g_i$ are the various form factors which are
functions of $q^2$. The number of independent form factors are
greatly reduced in the heavy quark symmetry limit. In this  limit,
the matrix elements of all the hadronic currents, irrespective of
their Dirac structure, can be given in terms of only two independent
form factors \cite{R46212} as \bea \langle \ll(p_\ll) | \bar s
\Gamma b | \lb (p_{\lb}) \rangle = \bar u_\ll [F_1(q^2) +\not\!{v}
F_2 (q^2) ] \Gamma u_{\lb}\;, \eea where $\Gamma$ is the product of
Dirac matrices, $v^\mu=p_{\lb}^\mu/ m_{\lb}$ is the four velocity of
$\Lambda_b$. These two sets of form facors are relalated to each
other as \bea
&& g_1=f_1=f_2^T=g_2^T=F_1 +\sqrt r F_2\;,\\
&& g_2=f_2=g_3=f_3=\frac{F_2}{m_{\lb}}\;,\\
&&g_3^T=\frac{F_2}{m_{\lb}}(m_{\lb}+m_\ll)\;,~~~~~~
f_3^T=-\frac{F_2}{m_{\lb}}(m_{\lb}-m_\ll)\\
&&f_1^T=g_1^T=\frac{F_2}{m_{\lb}}q^2\;, \eea where
$r=m_\ll^2/m_{\lb}^2$. Thus, using these form factors, the
transition amplitude can be written as
\bea \label{e3} \lefteqn{
{\cal M} = \frac{G \alpha}{4 \sqrt{2}\pi} V_{tb}V_{ts}^\ast \Bigg\{
\bar \ell \gamma^\mu \ell \, \bar u_\Lambda \Big[ A_1 \gamma_\mu
(1+\gamma_5) +
B_1 \gamma_\mu (1-\gamma_5) }\nnb \\
\ar i \sigma_{\mu\nu} q^\nu \big[ A_2 (1+\gamma_5) + B_2
(1-\gamma_5) \big] +q_\mu \big[ A_3 (1+\gamma_5) + B_3 (1-\gamma_5)
\big]\Big] u_{\Lambda_b}
\nnb \\
\ar \bar \ell \gamma^\mu \gamma_5 \ell \, \bar u_\Lambda \Big[E_1
\gamma_\mu (1-\gamma_5)  + i\sigma_{\mu\nu} q^\nu E_2 (1-\gamma_5) +
E_3 q^{\mu}(1-\gamma_5)\big] \Big] u_{\Lambda_b}\Bigg\}~, \eea where
$P=p_{\Lambda_b}+ p_\Lambda$. Explicit expressions of the functions
$A_i,~B_i,$ and $E_i$ $(i=1,2,3)$  are given as follows
\cite{R46212}: \bea A_1&=&-\frac{4 m_b}{m_{\Lambda_b}}~F_2~
C^{tot}_7\nnb\\A_2&=&-\frac{4 m_b}{q^2}~(F_1+\sqrt{r}F_2)~
C^{tot}_7\nnb\\A_3&=&-\frac{4 m_b
m_{\Lambda}}{q^2m_{\Lambda_b}}~F_2~
C^{tot}_7\nnb\\B_1&=&2(F_1+\sqrt{r}F_2)~C^{tot}_9\nnb\\B_2&=&\frac{2
F_2}{m_{\Lambda_b}}~C^{tot}_9\nnb\\B_3&=&\frac{4 m_b}{q^2}~F_2~
C^{tot}_7\nnb\\E_1&=&2(F_1+\sqrt{r}F_2)~
C^{tot}_{10}\nnb\\E_2&=&E_3=\frac{2F_2}{m_{\Lambda_b}}~C^{tot}_{10}\eea

From the expressions of the above-mentioned matrix elements Eq.
(\ref{e3}) we observe that $\Lambda_b \rar\Lambda \ell^+\ell^-$
decay is described in terms of many form factors. When HQET is
applied to the number of independent form factors, as it has already
been noted,  reduces to two ($F_1$ and $F_2$) irrelevant with the
Dirac structure of the corresponding operators and it is obtained in
\cite{R46210} that \bea \label{e4} \lla \Lambda(p_\Lambda) \vel \bar
s \Gamma b \ver \Lambda(p_{\Lambda_b}) \rra = \bar u_\Lambda
\Big[F_1(q^2) + \not\!v F_2(q^2)\Big] \Gamma u_{\Lambda_b}~, \eea
where $\Gamma$ is an arbitrary Dirac structure,
$v^\mu=p_{\Lambda_b}^\mu/m_{\Lambda_b}$ is the four--velocity of
$\Lambda_b$, and $q=p_{\Lambda_b}-p_\Lambda$ is the momentum
transfer. Comparing the general form of the form factors with
(\ref{e5}), one can easily obtain the following relations among them
(see also \cite{R4629}) \bea \label{e5}
g_1 \es f_1 = f_2^T= g_2^T = F_1 + \sqrt{r} F_2~, \nnb \\
g_2 \es f_2 = g_3 = f_3 = g_T^V = f_T^V = \frac{F_2}{m_{\Lambda_b}}~,\nnb \\
g_T^S \es f_T^S = 0 ~,\nnb \\
g_1^T \es f_1^T = \frac{F_2}{m_{\Lambda_b}} q^2~,\nnb \\
g_3^T \es \frac{F_2}{m_{\Lambda_b}} \ga m_{\Lambda_b} + m_\Lambda \dr~,\nnb \\
f_3^T \es - \frac{F_2}{m_{\Lambda_b}} \ga m_{\Lambda_b} - m_\Lambda
\dr~, \eea where $r=m_\Lambda^2/m_{\Lambda_b}^2$.

The differential decay rate of the $\Lambda_b \rar \Lambda \ell^+
\ell^-$ decay for any spin direction  can be written as: \bea
\label{e9} \ga \frac{d \Gamma}{ds}\dr_0 = \frac{G^2 \alpha^2}{192
\pi^5} \vel V_{tb} V_{ts}^\ast \ver^2 \lambda^{1/2}(1,r,s) v
\Big[{\cal T}_0(s) +\frac{1}{3} {\cal T}_2(s) \Big]~, \eea where
$\lambda(1,r,s) = 1 + r^2 + s^2 - 2 r - 2 s - 2 rs$ is the
triangle function and $v=\sqrt{1-4m_\ell^2/q^2}$ is the lepton
velocity. The explicit expressions for ${\cal T}_0$ and ${\cal
T}_2$ are given by: \bea {\cal T}_0&=&4m_{\Lambda_b}^2\Big\{
8m_{\ell}^2
m_{\Lambda_b}^2\hat{s}(1+r-\hat{s})|E_3|^2+16m_{\ell}^2
m_{\Lambda_b}\sqrt{r}(1-r+\hat{s})\mbox{\rm Re}[E_1^\ast
E_3]+\nnb\\&&8(2m_{\ell}^2+
m_{\Lambda_b}^2\hat{s})\{(1-r+\hat{s})m_{\Lambda_b}\sqrt{r}\mbox{\rm
Re}[A_1^\ast A_2+B_1^\ast B_2]
-\nnb\\&&m_{\Lambda_b}(1-r-\hat{s})\mbox{\rm Re}[A_1^\ast
B_2+A_2^\ast B_1]-2\sqrt{r}(\mbox{\rm Re}[A_1^\ast
B_1]+m_{\Lambda_b}^2\hat{s}\mbox{\rm Re}[A_2^\ast B_2])
\}+\nnb\\&&2\Big(
4m_{\ell}^2(1+r-\hat{s})+m_{\Lambda_b}^2\Big[(1-r)^2-\hat{s}^2\Big]\Big)\Big(|A_1|^2+|B_1|^2\Big)+\nnb\\&&
2m_{\Lambda_b}^2\Big(4m_{\ell}^2\Big[\lambda+(1+r-\hat{s})\hat{s}\Big]+
m_{\Lambda_b}^2\hat{s}\Big[(1-r)^2-\hat{s}^2\Big]\Big)\Big(|A_2|^2+|B_2|^2
\Big)-\nnb\\&&2\Big(4m_{\ell}^2(1+r-\hat{s})-m_{\Lambda_b}^2\Big[(1-r)^2-\hat{s}^2\Big]\Big)|E_1|^2+\nnb\\&&
2 m_{\Lambda_b}^3\hat{s}v^2\Big(4(1-r+\hat{s})\sqrt{r}\mbox{\rm
Re}[E_1^\ast
E_2]-m_{\Lambda_b}\Big[(1-r)^2-\hat{s}^2\Big]|E_2|^2\Big)\Big\}
\eea \bea {\cal
T}_2=-8m_{\Lambda_b}^4v^2\lambda\Big(|A_1|^2+|B_1|^2+|E_1|^2-m_{\Lambda_b}^2\hat{s}(|A_2|^2+|B_2|^2+|E_2|^2)\Big).\eea
\section{Double--lepton polarization asymmetries in the $\Lambda_b
\rar\Lambda \ell^+\ell^-$ decay}

In order to calculate the polarization asymmetries of both leptons
defined in the effective four fermion interaction of Eq.(\ref{e3}),
we must first define the orthogonal vectors $S$ in the rest frame of
$\ell^-$ and $W$ in the rest frame of $\ell^+$ (where these vectors
are the polarization vectors of the leptons). Note that we shall use
the subscripts $L$, $N$ and $T$ to correspond to the leptons being
polarized along the longitudinal, normal and transverse directions
respectively \cite{Kruger:1996cv,Bensalem:2002ni}.
\begin{eqnarray}
S^\mu_L & \equiv & (0, \mathbf{e}_{L}) ~=~ \left(0,
\frac{\mathbf{p}_-}{|\mathbf{p}_-|}
\right) , \nonumber \\
S^\mu_N & \equiv & (0, \mathbf{e}_{N}) ~=~ \left(0,
\frac{\mathbf{p_{\Lambda}} \times
\mathbf{p}_-}{|\mathbf{p_{\Lambda}} \times
\mathbf{p}_- |}\right) , \nonumber \\
S^\mu_T & \equiv & (0, \mathbf{e}_{T}) ~=~ \left(0, \mathbf{e}_{N}
\times \mathbf{e}_{L}\right) , \label{sec3:eq:1} \\
W^\mu_L & \equiv & (0, \mathbf{w}_{L}) ~=~ \left(0,
\frac{\mathbf{p}_+}{|\mathbf{p}_+|} \right) , \nonumber \\
W^\mu_N & \equiv & (0, \mathbf{w}_{N}) ~=~ \left(0,
\frac{\mathbf{p_{\Lambda}} \times
\mathbf{p}_+}{|\mathbf{p_{\Lambda}} \times
\mathbf{p}_+ |} \right) , \nonumber \\
W^\mu_T & \equiv & (0, \mathbf{w}_{T}) ~=~ (0, \mathbf{w}_{N} \times
\mathbf{w}_{L}) , \label{sec3:eq:2}
\end{eqnarray}
where $\mathbf{p}_+$, $\mathbf{p}_-$ and $\mathbf{p_{\Lambda}}$ are
the three momenta of the $\ell^+$, $\ell^-$ and $\Lambda$ particles
respectively. On boosting the vectors defined by
Eqs.(\ref{sec3:eq:1},\ref{sec3:eq:2}) to the c.m. frame of the
$\ell^- \ell^+$ system only the longitudinal vector will be boosted,
whilst the other two vectors remain unchanged. The longitudinal
vectors after the boost will become;
\begin{eqnarray}
S^\mu_L & = & \left( \frac{|\mathbf{p}_-|}{m_\ell}, \frac{E_{\ell}
\mathbf{p}_-}{m_\ell |\mathbf{p}_-|} \right) , \nonumber \\
W^\mu_L & = & \left( \frac{|\mathbf{p}_-|}{m_\ell}, - \frac{E_{\ell}
\mathbf{p}_-}{m_\ell |\mathbf{p}_-|} \right) . \label{sec3:eq:3}
\end{eqnarray}
The polarization asymmetries can now be calculated using the spin
projector ${1 \over 2}(1 + \gamma_5 \!\!\not\!\! S)$ for $\ell^-$
and the spin projector ${1 \over 2}(1 + \gamma_5\! \not\!\! W)$ for
$\ell^+$.

\par Equipped with the above expressions we now define the various
 double lepton polarization asymmetries. The double
lepton polarization asymmetries are defined as
\cite{Bensalem:2002ni};
\begin{eqnarray}
{\cal P}_{xy} & \equiv & \frac{\left( \frac{d\Gamma( S_x, W_y
)}{d\hat{s}} - \frac{d\Gamma( - S_x, W_y )}{d\hat{s}} \right) -
\left( \frac{d\Gamma( S_x, - W_y )}{d\hat{s}} - \frac{d\Gamma(- S_x,
- W_y )}{d\hat{s}}\right)} {\left( \frac{d\Gamma( S_x, W_y
)}{d\hat{s}} + \frac{d\Gamma( - S_x, W_y )}{d\hat{s}} \right) +
\left( \frac{d\Gamma( S_x, - W_y )}{d\hat{s}} + \frac{d\Gamma( -
S_x, - W_y )}{d\hat{s}}\right)} , \label{sec3:eq:5}
\end{eqnarray}
where the sub-indices $x$ and $y$ can be either $L$, $N$ or $T$. And
the double polarization asymmetries are; \bea\label{pll}
P_{LL}&=&\frac{8
m^4_{\lb}}{3\Delta}Re\Bigg\{12\hml(1-\sqrt{\rr})(1+2\sqrt{\rr}+\rr-\ss)E_1
F_2^\ast \\ \nnb &+&3\ss(1+2\sqrt{\rr}+\rr-\ss)[v^2
|F_1|^2+|F_2|^2+4 m_{\lb}\hml F_3 F_2^\ast ]\\ \nnb&-&12
m_{\lb}\sqrt{\rr}(1-\rr+\ss) [\ss(1+v^2)(A_1 A_2^\ast+B_1
B_2^\ast)-4 \hml^2 E_1 E_3^\ast ]
 \\ \nnb &+&12 m_{\lb}(1-\rr-\ss)[\ss(1+v^2)(A_1 B_2^\ast+A_2 B_1^\ast)]
 \\ \nnb&+&24\sqrt{\rr}\ss(1+v^2)(A_1 B_1^\ast+m_{\lb}^2\ss A_2 B_2^\ast)
+24 m_{\lb}^2 \hml^2 \ss(1+\rr-\ss)|E_3|^2\\ \nnb
&-&2(1+v^2)[1+\rr^2-\rr(2-\ss)+\ss(1-2\ss)](|A_1|^2+|B_1|^2) \\
\nnb &-&2
[(5v^2-3)(1-\rr)^2+4\hml^2(1+\rr)+2\ss(1+8\hml^2+\rr)-4\ss^2]|E_1|^2
\\ \nnb &-&2m_{\lb}^2(1+v^2)\ss [2+2\rr^2-\ss(1+\ss)-\rr(4+\ss)](|A_2|^2+|B_2|^2)
\\ \nnb &-& 4 m_{\lb}^2\ss v^2 [2(1+\rr^2)-\ss(1+\ss)-\rr(4+\ss)]|E_2|^2
\\ \nnb &-&24 m_{\lb}\sqrt{\rr}\ss(1-\rr+\ss)v^2E_1 E_2^\ast\Bigg\}\eea

\bea\label{pnl}P_{NL}=-P_{LN}&=&\frac{4 \pi
m^4_{\lb}\sqrt{\lambda}}{\Delta\sqrt{\ss}}Im\Bigg\{
4\hml(1-\rr)B_1^\ast E_1 + 4 m_{\lb}\hml \ss (A_1^\ast
E_3-A_2^\ast E_1)\\ \nnb&+&\ss(1+\sqrt{\rr})(A_1+B_1)^\ast
F_2+4m_{\lb}\hml\sqrt{\rr}\ss (B_1^\ast E_3+B_2^\ast E_1)\\ \nnb
&-&m_{\lb}\ss^2[B_2^\ast(F_2+4m_{\lb}\hml E_3)+A_2^\ast F_2]-\ss
v^2 [E_1 F_1^\ast-\sqrt{\rr}E_1^\ast F_1]+m_{\lb}\ss^2 v^2
F_1E_2^\ast \Bigg\}\eea

 \bea\label{plt}P_{TL}=P_{LT}&=& \frac{4
\pi m_{\lb}^4\sqrt{\lambda}v}{\Delta\sqrt{\ss}}
 Re\Bigg\{4\hml(1-\rr)|E_1|^2-4\hml ~\ss~ B_1 E_1^\ast \\ \nnb&-& 4 \hml~ \ss~ m_{\lb}
 (A_2 E_1^\ast-A_1 E_2^\ast)-\ss (1+\sqrt{\rr})[(A_1+B_1)F_1^\ast-E_1 F_2^\ast]
 \\ \nnb &-&4 m_{\lb}^2 \ss (1-\rr)\hml B_2 E_2^\ast-4m_{\lb} \hml \sqrt{\rr}\ss
 [B_1 E_2^\ast+(B_2 -E_2-E_3)E_1^\ast]\\ \nnb&+& m_{\lb} \ss^2 [(A_2+B_2)F_1^\ast-
 E_2 F_2^\ast-4m_{\lb}\hml E_2 E_3^\ast]\Bigg\} \eea

\bea\label{pnt}P_{NT}&=& \frac{16 m_{\lb}^4 v}{3\Delta}Im\Bigg\{4
\lambda B_1 E_1^\ast -6 \hml
(1-\sqrt{\rr})(1+2\sqrt{\rr}+\rr-\ss)E_1 F_1^\ast\\ \nnb &+&4
m_{\lb}^2 \lambda \ss E_2B_2^\ast-3\ss(1+2\sqrt{\rr}+\rr-\ss)(F_2+2
m_{\lb}\hml E_3)F_1^\ast \Bigg\}\eea
 \bea\label{ptn}P_{TN}&=&
-\frac{16 m_{\lb}^4 v}{3\Delta}Im\Bigg\{4 \lambda B_1 E_1^\ast +6
\hml (1-\sqrt{\rr})(1+2\sqrt{\rr}+\rr-\ss)E_1 F_1^\ast\\ \nnb &+&4
m_{\lb}^2 \lambda \ss
E_2B_2^\ast+3\ss(1+2\sqrt{\rr}+\rr-\ss)(F_2+2 m_{\lb}\hml
E_3)F_1^\ast \Bigg\}\eea

\bea\label{pnn}
 P_{NN}&=&\frac{8
m^4_{\lb}}{3\hat{s}\Delta}Re\Bigg\{96\hat{m}_l^2\sqrt{\hat{r_\Lambda}}\hat{s}A_1B_1^\ast
\\ \nnb &-&
48m_{\lb}\hat{m}_l^2\sqrt{\hat{r_\Lambda}}\hat{s}(1-\hat{r}_\Lambda+\hat{s})(A_1A_2^\ast+B_1B_2^\ast) \\
\nnb &+&
12\hat{m}_l\hat{s}(1-\sqrt{\hat{r}_\Lambda})(1+2\sqrt{\hat{r}_\Lambda}+\hat{r}_\Lambda-\hat{s})E_1F_2^\ast
\\ \nnb
&+&
3\hat{s}^2(1+2\sqrt{\hat{r}_\Lambda}+\hat{r}_\Lambda-\hat{s})[|F_2|^2+4m_{\Lambda_{b}}\hat{m}_lE_3F_2^\ast]
\\ \nnb &+&
24m_{\lb}\hat{m}_l^2\hat{s}[m_{\lb}\hat{s}(1+\hat{r}_\Lambda-\hat{s})|E_3|^2
+2\sqrt{\hat{r}_\Lambda}(1-\hat{r}_\Lambda+\hat{s})E_1E_3^\ast]
\\ \nnb
&+&48m_{\lb}\hat{m}_l^2\hat{s}(1-\hat{r}_\Lambda-\hat{s})(A_1B_2^\ast+A_2B_1^\ast)
\\ \nnb &-& 4[\lambda\hat{s}+2\hat{m}_l^2(1+\hat{r}_\Lambda^2-2\hat{r}_\Lambda+\hat{r}_\Lambda\hat{s}+\hat{s}-2\hat{s}^2)]
(|A_1|^2+|B_1|^2-|E_1|^2)\\ \nnb
 &+&
 96m_{\lb}^2\hat{m}_l^2\sqrt{\hat{r_\Lambda}}\hat{s}^2A_2B_2^\ast
 -4m_{\lb}^2\lambda\hat{s}^2v^2|E_2|^2 \\ \nnb
 &+&
 4m_{\lb}^2\hat{s}\{\lambda\hat{s}-2\hat{m}_l^2[2(1+\hat{r}^2)-\hat{s}(1+\hat{s})-\hat{r}(4+\hat{s})]\}(|A_2|^2+|B_2|^2)
 \\ \nnb &-&
 3\hat{s}^2v^2(1+2\sqrt{\hat{r_\Lambda}}+\hat{r_\Lambda}-\hat{s})|F_1|^2\Bigg\}
 \eea

\bea\label{ptt}
P_{TT}&=&\frac{
8m^4_{\lb}}{3\hat{s}\Delta}Re\Bigg\{-96\hat{m}_l^2\sqrt{\hat{r_\Lambda}}\hat{s}A_1B_1^\ast
\\ \nnb &-&
48 m_{\lb}\hat{m}_l^2\sqrt{\hat{r_\Lambda}}\hat{s}(1-\hat{r}_\Lambda+\hat{s})E_1E_3^\ast \\
\nnb &-&
12\hat{m}_l\hat{s}(1-\sqrt{\hat{r}_\Lambda})(1+2\sqrt{\hat{r}_\Lambda}+\hat{r}_\Lambda-\hat{s})E_1F_2^\ast
\\ \nnb
&-&
96m_{\lb}^2\hat{m}_l^2\sqrt{\hat{r_\Lambda}}\hat{s}^2A_2B_2^\ast-
3\hat{s}^2(1+2\sqrt{\hat{r}_\Lambda}+\hat{r}_\Lambda-\hat{s})[|F_2|^2+4m_{\lb}\hat{m}_lE_3F_2\ast]\\
\nnb &-&
24m_{\lb}\hat{m}_l^2\hat{s}[m_{\lb}\hat{s}(1+\hat{r}_\Lambda-\hat{s})|E_3|^2
-2\sqrt{\hat{r}_\Lambda}(1-\hat{r}_\Lambda+\hat{s})(A_1A_2^\ast+B_1B_2^\ast)]
\\ \nnb &-& 48m_{\lb}\hat{m}_l^2\hat{s}(1-\hat{r}_\Lambda-\hat{s})(A_1B_2^\ast+A_2B_1^\ast)
\\ \nnb &-& 4[\lambda\hat{s}-2\hat{m}_l^2(1+\hat{r}_\Lambda^2-2\hat{r}_\Lambda+\hat{r}_\Lambda\hat{s}+\hat{s}-2\hat{s}^2)]
(|A_1|^2+|B_1|^2)\\ \nnb
 &+& 4m_{\lb}^2\hat{s}\{\lambda\hat{s}+\hat{m}_l^2[4(1-\hat{r}_\Lambda)^2-2\hat{s}
(1+\hat{r}_\Lambda)-2\hat{s}^2]\}(|A_2|^2+|B_2|^2)
 \\ \nnb &+&
 4\{\lambda\hat{s}-2\hat{m}_l^2[5(1-\hat{r}_\Lambda)^2-v\hat{s}(1+\hat{r}_\Lambda)+2\hat{s}^2]\}|E_1|^2
 \\ \nnb &-&4m_{\lb}^2\lambda\hat{s}^2v^2|E_2|^2+3\hat{s}^2v^2(1+2\sqrt{\hat{r}_\Lambda}+\hat{r}_\Lambda-\hat{s})|F_1|^2
 \Bigg\}
 \eea

\section{Numerical analysis}

In this section, we examine the dependence of the double lepton
polarizations to the fourth quark mass($m_{t'}$) and the product of
quark mixing matrix elements ($V_{t^\prime b}^\ast V_{t^\prime
s}=r_{sb}e^{i\phi_{sb}}$). For numerical evaluation we use the
various particle masses and lifetimes of $\lb$ baryon from
\cite{PDG}. The quark masses (in GeV) used are $m_b$=4.8,
$m_c$=1.35, the CKM matrix elements as $|V_{cb} V_{cs}^*|=0.041$,
$\alpha=1/128$ and the weak mixing angle $\sin^2 \theta_W=0.23$. For
the form factors we use the values calculated in the QCD sum rule
approach in combination with HQET \cite{R46210, R46214}, which
reduces the number of quite many form factors into two. The $s$
dependence of these form factors can be represented in the following
way \bea F(q^2) = \frac{F(0)}{\ds 1-a_F s + b_F s^2}~, \nnb \eea
where parameters $F_i(0),~a$ and $b$ are listed in table 2.
\begin{table}
\renewcommand{\arraystretch}{1.5}
\addtolength{\arraycolsep}{3pt}
$$
\begin{array}{|l|ccc|}
\hline & F(0) & a_F & b_F \\ \hline F_1 &
\phantom{-}0.462 & -0.0182 & -0.000176 \\
F_2 & -0.077 & -0.0685 &\phantom{-}0.00146 \\ \hline
\end{array}
$$
\caption{Transition form factors for $\Lambda_b \rar \Lambda \ell^+
\ell^-$ decay in the QCD sum rules method.}
\renewcommand{\arraystretch}{1}
\addtolength{\arraycolsep}{-3pt}
\end{table}

We use the next--to--leading order logarithmic approximation for the
resulting values of the Wilson coefficients $C_9^{eff},~C_7$ and
$C_{10}$ in the SM \cite{R46215,R46216} at the re--normalization
point $\mu=m_b$. It should be noted that, in addition to short
distance contribution, $C_9^{eff}$ receives also long distance
contributions from the real $\bar c c$ resonant states of the
$J/\psi$ family. In the present work, we do not take  the long
distance effects into account.
 In order to perform quantitative analysis of the
double lepton polarizations, the values of the new
parameters($m_{t'},\,r_{sb},\,\phi_{sb}$) are needed. Using the
experimental values of $B\rar X_s \gamma$ and $B\rar X_s \ell^+
\ell^-$, the bound on $r_{sb}\sim\{0.01-0.03\}$ has been obtained
\cite{Arhrib:2002md} for $\phi_{sb}\sim\{0-2\pi\}$ and
$m_{t'}\sim\{300,400\}~$(GeV). We do a different and somehow more
general analysis with the recent world average value\cite{NNLL} of
the \bea {\cal{B}}(B\rar X_s \ell^+ \ell^-)=(1.6\pm 0.51)\times
10^{-6},\,\,\,\,\, \ell=(\mu,\, e)\eea in the low dilepton invariant
mass region($1\texttt{GeV}^2<q^2<6\texttt{GeV}^2$). We chose
$r_{sb}\sim\{0.005-0.03\}$ , $ \phi_{sb}\sim\{60^\circ-120^\circ\}$
and $ 1\sigma $ level deviation from the experimental value, then we
obtain the constrain on $m_{t'}$ (see Table 3, 4 and 5 ). However,
the most general analysis about range
 of new parameters ($\phi_{sb},~ r_{sb}, m_{t'}$) considering the recent experimental value
 of  $B\rar X_s \ell^+ \ell^-$ are still incomplete in some sense.
 We plan to do that in our next work. In the
foregoing numerical analysis, we vary $m_{t'}$ in the range $175\le
m_{t'} \le 600$GeV. The lower range is because of the fact that the
fourth generation up quark should be heavier than the third
ones($m_t \leq m_{t'}$)\cite{Sultansoy:2000dm}. The upper range
comes from the experimental bounds on the $\rho$ and $S$ parameters
of SM, which we mentioned above(see Introduction). At the same time
we will show the constrain on $m_{t'}$ coming from the experimental
values of the $B\rar X_s \ell^+ \ell^-$ in our figures.

Before performing numerical analysis, few words about lepton
polarizations are in order. From explicit expressions of the
lepton polarizations one can easily see that they depend on both
$\ss$ and the new parameters($m_{t'},\,r_{sb}$). We should
eliminate the dependence of the lepton polarization on one of the
variables. We eliminate the variable $\hat{s}$ by performing
integration over $\ss$ in the allowed kinematical region. The
total branching ratio and the averaged lepton polarizations are
defined as \bea {\cal B}_r&=&\ds \int_{4
m_\ell^2/m_{\Lambda_b}^2}^{(1-\sqrt{r})^2}
 \frac{d{\cal B}}{ds} ds,
\nnb\\\lla P_{ij} \rra &=& \frac{\ds \int_{4
m_\ell^2/m_{\Lambda_b}^2}^{(1-\sqrt{r})^2} P_{ij} \frac{d{\cal
B}}{ds} ds} {{\cal{B}}_r}~. \eea

We ignore to show $\lla P_{LL} \rra$  for the $\Lambda_b \rar
\Lambda \mu^+ \mu^-$ decay, since its value is quite small.

Figs. (1)--(33) show the dependency of $\lla P_{ij} \rra$ for the
$\Lambda_b \rar \Lambda \ell^+ \ell^-$ decay at four values of
$r_{sb}:0.005, 0.01, 0.02, 0.03$ on the $ m_{t'} $ for $\ell=\mu,\,
\tau$ channels  at three different values of $\phi_{sb}:60^\circ,
90^\circ 120^\circ$. The $\bullet \!\!\!\backslash$ sign in figures
show the experimental upper limit on $ m_{t'} $ coming from the
$B\rar X_s \ell^+ \ell^-$ analysis. From these figures, we obtain
the following results.
\begin{table}
\renewcommand{\arraystretch}{1.5}
\addtolength{\arraycolsep}{3pt}
$$
\begin{array}{|c|c|c|c |c|}
\hline  r_{sb} & 0.005 & 0.01 &0.02 & 0.03 \\
\hline
m_{t'}(GeV) & 739 &529 & 385 & 331\\
\hline
\end{array}
$$
\caption{The extracted maximum experimental limit of $ m_{t'} $ for
$\phi_{sb}=\pi/3$}
\renewcommand{\arraystretch}{1}
\addtolength{\arraycolsep}{-3pt}
\end{table}

\begin{table}
\renewcommand{\arraystretch}{1.5}
\addtolength{\arraycolsep}{3pt}
$$
\begin{array}{|c|c|c|c |c|}
\hline  r_{sb} & 0.005 & 0.01 &0.02 & 0.03 \\
\hline
m_{t'}(GeV) &511 &373 & 289 & 253\\
\hline
\end{array}
$$
\caption{The extracted maximum experimental limit of $ m_{t'} $ for
$\phi_{sb}=\pi/2$}
\renewcommand{\arraystretch}{1}
\addtolength{\arraycolsep}{-3pt}
\end{table}

\begin{table}
\renewcommand{\arraystretch}{1.5}
\addtolength{\arraycolsep}{3pt}
$$
\begin{array}{|c|c|c|c |c|}
\hline  r_{sb} & 0.005 & 0.01 &0.02 & 0.03 \\
\hline
m_{t'}(GeV) &361 &283 & 235 & 217\\
\hline
\end{array}
$$
\caption{The extracted maximum experimental limit of $ m_{t'} $ for
$\phi_{sb}=2\pi/3$}
\renewcommand{\arraystretch}{1}
\addtolength{\arraycolsep}{-3pt}
\end{table}

\begin{itemize}
\item $\lla P_{LL}\rra$ for $\Lambda_b \rar \Lambda \tau^+ \tau^-$
decay depends strongly on the SM4 parameters. Firstly, there exist
regions of the $ m_{t'} $ where $\lla P_{LL}\rra$ departs
considerably from the SM3 result. Secondly, there is an
experimentally allowed regions for the value of $ m_{t'} $ where
$\lla P_{LL}\rra$ changes its sign(see Fig. 1)in compare with SM3
value. The measurement of the sign of $\lla P_{LL}\rra$ for $\tau$
channel can be used as a good tool to look for new Physics effects.
More precisely, the results can be used to indirect search to look
for fourth generation of quarks.

\item $\lla P_{LN}\rra$($\lla P_{NL}\rra$)and $\lla
P_{TN}\rra$(-$\lla P_{NT}\rra$)  show strong dependency on SM4
parameters. It is increasing function in the experimentally
allowed regions and decreasing function of $\phi_{sb}$ for both
$\mu$ and $\tau$ channels(see Figs. 4--9 and Figs. 22--27). The
SM3 values of $\lla P_{LN}\rra$($\lla P_{NL}\rra$)and $\lla
P_{TN}\rra$(-$\lla P_{NT}\rra$) approximately vanish for both .
But, it receive the maximum values of $\approx 0.1$ , $\approx
0.4$ and $\approx 0.4$ and minimum value $\approx -0.1$ for $\mu$
and $\tau $ channels(see Figs. 4, 5 and Figs. 22, 25),
respectively. The results can be used to look for NP. It should be
noted that, when $ m_{t'}\rar m_t $ the SM4 result could coincide
with the SM3 (see Strategy). The deviation of $\approx 1\%$ from
the SM3 values for $\mu$ channel(see Figs. 4, 6, 8 and 25--27) is
because, firstly, we use the NNLL calculation for the SM3 values
of Wilson coefficients($C_i^{SM}$) and LL formulas for
$C_i^{new}$(see Eq. (\ref{c4})). Secondly, our numerical
integration for $P_{ij}$ has the same order of error.

\item $\lla P_{LT}\rra$($\lla P_{TL}\rra$) are very sensitive to
$m_{t'}$ and $r_{sb}$ and less sensitive to the $\phi_{sb}$. We
observe that $\lla P_{LT}\rra$($\lla P_{TL}\rra$) exceeds the SM3
prediction 2 and 3 times for $\mu$ and $\tau$ channels(see Figs.
10--15), respectively. Such behaviors can serve as a good test for
establishing new physics beyond the SM.

\item $\lla P_{TT}\rra$ and $\lla P_{NN}\rra$ are quite sensitive
to the existence of the SM4 parameters either in experimentally
allowed regions or in the hole region. They are increasing and
decreasing function of $m_{t'}$ if $\phi_{sb}=60^\circ$. But, they
are just increasing for $\mu$ case and decreasing for $\tau$ case
if $\phi_{sb}=90^\circ$ or $\phi_{sb}=120^\circ$. In the presence
of the 4th generation, the magnitude of $\lla P_{TT}\rra$ can
exceed the SM result 8 and 2 times and $\lla P_{NN}\rra$ can
exceed the SM result 4 and 5 times for $\mu$ and $\tau$ cases,
respectively. Therefore, determination of the magnitude of $\lla
P_{TT}\rra$ and $\lla P_{NN}\rra$ can give unambiguous information
about the existence of the new generation.
\end{itemize}
At the end of this section, let us discuss the problem of
measurement of the lepton polarization asymmetries in experiments.
Experimentally, to measure an asymmetry $\la P_{ij} \ra$ of the
decay with the branching ratio ${\cal B}$ at $n \sigma$ level, the
required number of events (i.e., the number of $B \bar{B}$ pair)
are given by the expression \bea N = \frac{n^2}{{\cal B} s_1 s_2
\la P_{ij} \ra^2}~,\nnb \eea where $s_1$ and $s_2$ are the
efficiencies of the leptons. Typical values of the efficiencies of
the $\tau$--leptons range from $50\%$ to $90\%$ for their various
decay modes (see for example \cite{R6625} and references therein),
and the error in $\tau$--lepton polarization is estimated to be
about $(10 -15)\%$ \cite{R6626}. As a result, the error in
measurement of the $\tau$--lepton asymmetries is of the order of
$(20 - 30)\%$, and the error in obtaining the number of events is
about $50\%$.

From the expression for $N$ we see that, in order to observe the
lepton polarization asymmetries in $\Lambda_b \rar\Lambda \mu^+
\mu^-$ and $\Lambda_b \rar\Lambda \tau^+ \tau^-$ decays at
$3\sigma$ level, the minimum number of required events are (for
the efficiency of $\tau$--lepton we take $0.5$):

\begin{itemize}
\item for the $\Lambda_b \rar\Lambda \mu^+ \mu^-$ decay \bea N =
\left\{ \begin{array}{ll}
2.0 \times 10^{6}  & (\mbox{\rm for} \lla P_{LL} \rra)~,\\
2.0 \times 10^{8}  & (\mbox{\rm for} \lla P_{LT} \rra=\lla P_{TL}
\rra, \lla P_{NN} \rra,\lla P_{TT} \rra)~,\end{array} \right. \nnb
\eea

\item for $\Lambda_b \rar\Lambda \tau^+ \tau^-$ decay \bea N =
\left\{ \begin{array}{ll} (4.0 \pm 2) \times 10^{9}  & (\mbox{\rm
for} \lla P_{LT}\rra,
\lla P_{NN} \rra)~,\\
(1.0 \pm 0.5) \times 10^{9}  & (\mbox{\rm for} \lla P_{TT} \rra)~,\\
(2.0 \pm 1.0) \times 10^{11} & (\mbox{\rm for} \lla P_{LN} \rra,
\lla P_{NL} \rra)~,\\
(9.0 \pm 4.5) \times 10^{8} & (\mbox{\rm for} \lla P_{TL} \rra)~.
\end{array} \right.
\nnb \eea
\end{itemize}

The number of $B \bar{B}$ pairs, that are produced at B--factories
and LHC \, are about $\sim 5\times 10^8$ and $10^{12}$,
respectively. As a result of a comparison of these numbers and
$N$, we conclude that, only $\lla P_{LL} \rra$ in the $\Lambda_b
\rar\Lambda \mu^+ \mu^-$ decay and  $\lla P_{LT} \rra$, $\lla
P_{NN} \rra$ and $\lla P_{TL} \rra$ in the $\Lambda_b \rar\Lambda
\tau^+ \tau^-$ decay, can be detectable at LHC.

To sum up, we presented the most general analysis of the
double--lepton polarization asymmetries in the $\Lambda_b
\rar\Lambda \ell^+ \ell^-$ decay using the SM with fourth generation
in this study. We also studied the dependence of the averaged
double--lepton polarization asymmetries on the SM4 parameters. Our
results showed that the averaged double--lepton polarization
asymmetries are strongly dependent on the fourth quark ($m_{t'}$)
and the product of quark mixing matrix elements ($V_{t^\prime
b}^\ast V_{t^\prime s}=r_{sb}e^{i\phi_{sb}}$). Thus, the
experimental determination of  both the sign and the magnitude of
the $\lla P_{ij} \rra$ can serve as a good tool to look for new
physics beyond the SM. More precisely, the study of the averaged
double--lepton polarization asymmetries can serve as good tool for
searching new generation of quarks.

\section{Acknowledgment}
The authors would like to thank T. M. Aliev for his useful
discussions. Also, the authors would like to thank TWAS, Iranian
chapter via ISOMO, for their partially support.

\newpage

\section*{Figure captions}
{\bf Fig. (1)} The dependence of $\lla P_{LL}\rra$ for $\Lambda_b
\rar \Lambda \tau^+ \tau^-$decay on $m_{t'}$, at four fixed values
of $r_{sb}:0.005, 0.01, 0.02, 0.03$ and
$\phi_{sb}=60^\circ$ .\\ \\
{\bf Fig. (2)} The same as in Fig. (1), but for $\phi_{sb}=90^\circ$.\\ \\
{\bf Fig. (3)} The same as in Fig. (1), but for $\phi_{sb}=120^\circ$.\\ \\
{\bf Fig. (4)} The dependence of $\lla P_{LN}\rra$ for $\Lambda_b
\rar \Lambda \mu^+ \mu^-$decay on $m_{t'}$, at four  fixed values of
$r_{sb}:0.005, 0.01, 0.02, 0.03$ and
$\phi_{sb}=60^\circ$ .\\ \\
{\bf Fig. (5)} The same as in Fig. (4), but for the $\tau$ lepton.\\\\
{\bf Fig. (6)} The same as in Fig. (4), but for $\phi_{sb}=90^\circ$.\\ \\
{\bf Fig. (7)} The same as in Fig. (6), but for the $\tau$ lepton.\\\\
{\bf Fig. (8)} The same as in Fig. (4), but for $\phi_{sb}=120^\circ$.\\ \\
{\bf Fig. (9)} The same as in Fig. (8), but for the $\tau$ lepton.\\\\
{\bf Fig. (10)} The dependence of $\lla P_{LT}\rra$ for $\Lambda_b
\rar \Lambda \mu^+ \mu^-$decay on $m_{t'}$, at four  fixed values of
$r_{sb}:0.005, 0.01, 0.02, 0.03$ and
$\phi_{sb}=60^\circ$ .\\ \\
{\bf Fig. (11)} The same as in Fig. (10), but for the $\tau$ lepton.\\\\
{\bf Fig. (12)} The same as in Fig. (10), but for $\phi_{sb}=90^\circ$.\\ \\
{\bf Fig. (13)} The same as in Fig. (12), but for the $\tau$ lepton.\\\\
{\bf Fig. (14)} The same as in Fig. (10), but for $\phi_{sb}=120^\circ$.\\ \\
{\bf Fig. (15)} The same as in Fig. (14), but for the $\tau$ lepton.\\\\
{\bf Fig. (16)} The dependence of $\lla P_{TT}\rra$ for $\Lambda_b
\rar \Lambda \mu^+ \mu^-$decay on $m_{t'}$, at four fixed values of
$r_{sb}:0.005, 0.01, 0.02, 0.03$ and
$\phi_{sb}=60^\circ$ .\\ \\
{\bf Fig. (17)} The same as in Fig. (16), but for the $\tau$ lepton.\\\\
{\bf Fig. (18)} The same as in Fig. (16), but for $\phi_{sb}=90^\circ$.\\ \\
{\bf Fig. (19)} The same as in Fig. (18), but for the $\tau$ lepton.\\\\
{\bf Fig. (20)} The same as in Fig. (16), but for $\phi_{sb}=120^\circ$.\\ \\
{\bf Fig. (21)} The same as in Fig. (20), but for the $\tau$ lepton.\\\\
{\bf Fig. (22)} The dependence of $\lla P_{TN}\rra$ for $\Lambda_b
\rar \Lambda \mu^+ \mu^-$decay on $m_{t'}$, at four fixed values of
$r_{sb}:0.005, 0.01, 0.02, 0.03$ and
$\phi_{sb}=60^\circ$ .\\ \\
{\bf Fig. (23)} The same as in Fig. (22), but for $\phi_{sb}=90^\circ$.\\ \\
{\bf Fig. (24)} The same as in Fig. (22), but for $\phi_{sb}=120^\circ$.\\ \\
{\bf Fig. (25)} The dependence of $\lla P_{NT}\rra$ for $\Lambda_b
\rar \Lambda \tau^+ \tau^-$decay on $m_{t'}$, at four  fixed values
of $r_{sb}:0.005, 0.01, 0.02, 0.03$ and
$\phi_{sb}=60^\circ$ .\\ \\
{\bf Fig. (26)} The same as in Fig. (25), but for $\phi_{sb}=90^\circ$.\\ \\
{\bf Fig. (27)} The same as in Fig. (25), but for $\phi_{sb}=120^\circ$.\\ \\
{\bf Fig. (28)} The dependence of $\lla P_{NN}\rra$ for $\Lambda_b
\rar \Lambda \mu^+ \mu^-$decay on $m_{t'}$, at four fixed values of
$r_{sb}:0.005, 0.01, 0.02, 0.03$ and
$\phi_{sb}=60^\circ$ .\\ \\
{\bf Fig. (29)} The same as in Fig. (28), but for the $\tau$ lepton.\\\\
{\bf Fig. (30)} The same as in Fig. (28), but for $\phi_{sb}=90^\circ$.\\ \\
{\bf Fig. (31)} The same as in Fig. (30), but for the $\tau$ lepton.\\\\
{\bf Fig. (32)} The same as in Fig. (28), but for $\phi_{sb}=120^\circ$.\\ \\
{\bf Fig. (33)} The same as in Fig. (32), but for the $\tau$ lepton.\\\\

\newpage

\begin{figure}
\vskip 2.5 cm
    \includegraphics{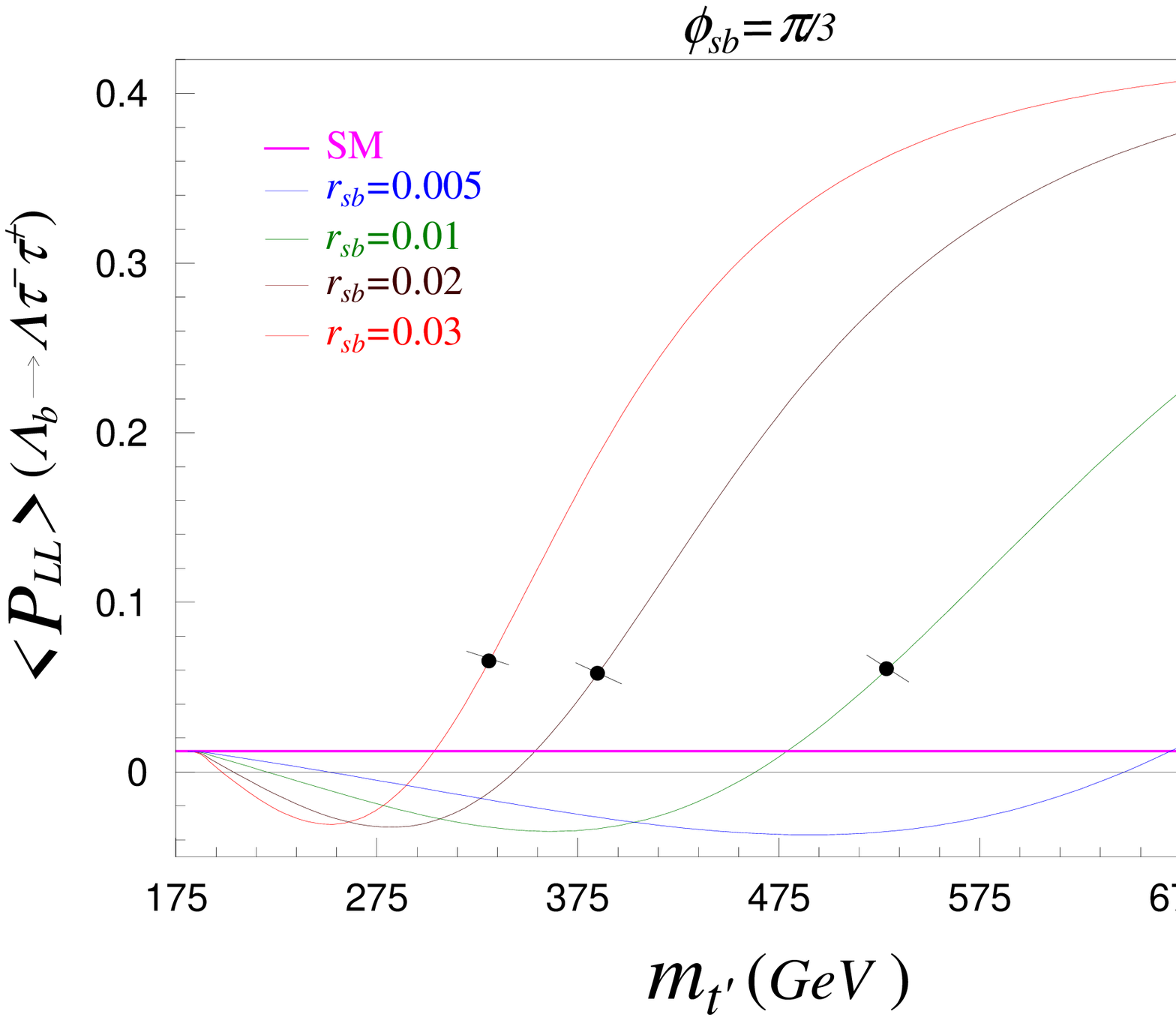}
\vskip 7.8cm \caption{}
\end{figure}

\begin{figure}
\vskip 2.5 cm
    \includegraphics{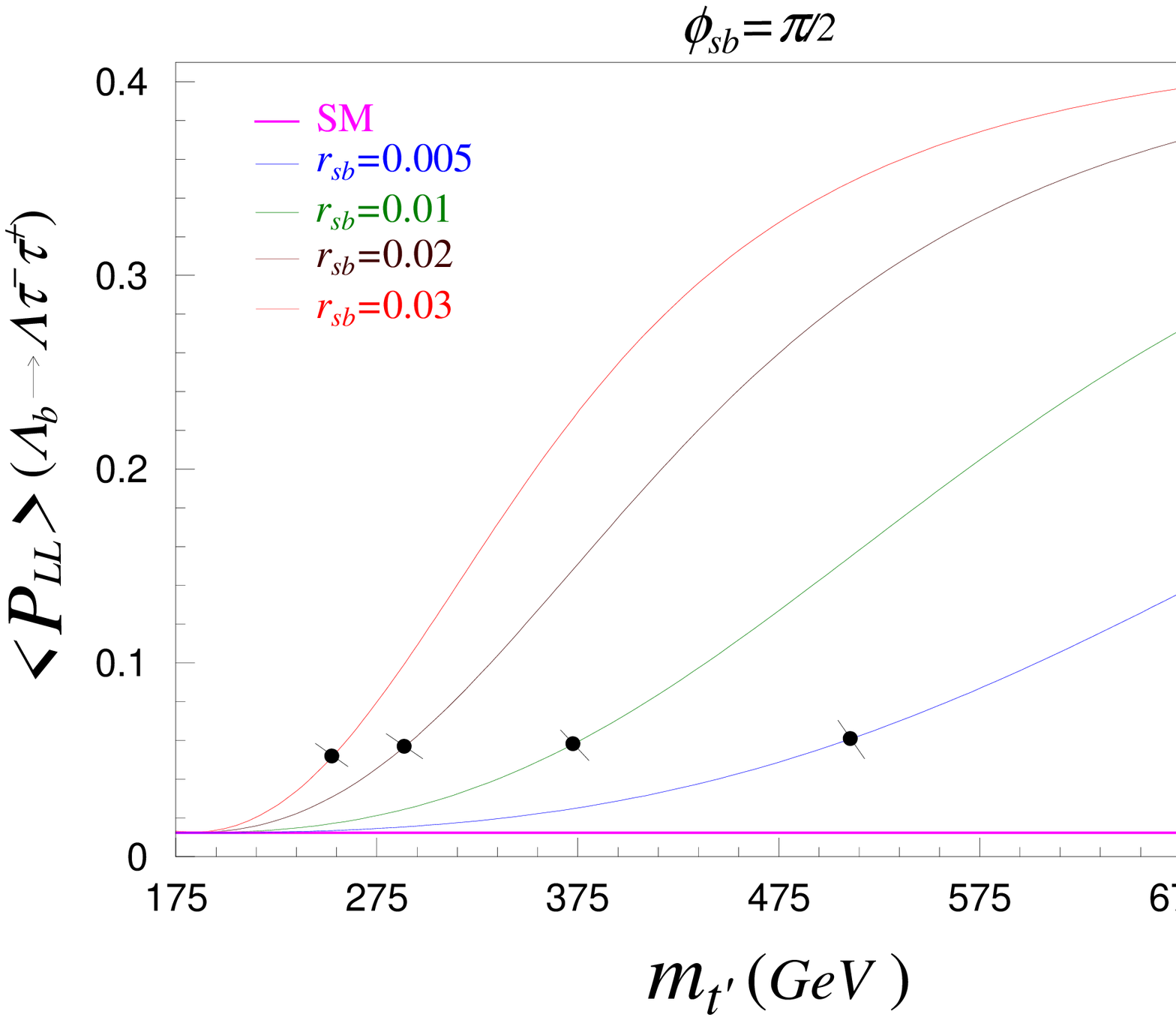}
\vskip 7.8cm \caption{}
\end{figure}

\begin{figure}
\vskip 2.5 cm
    \includegraphics{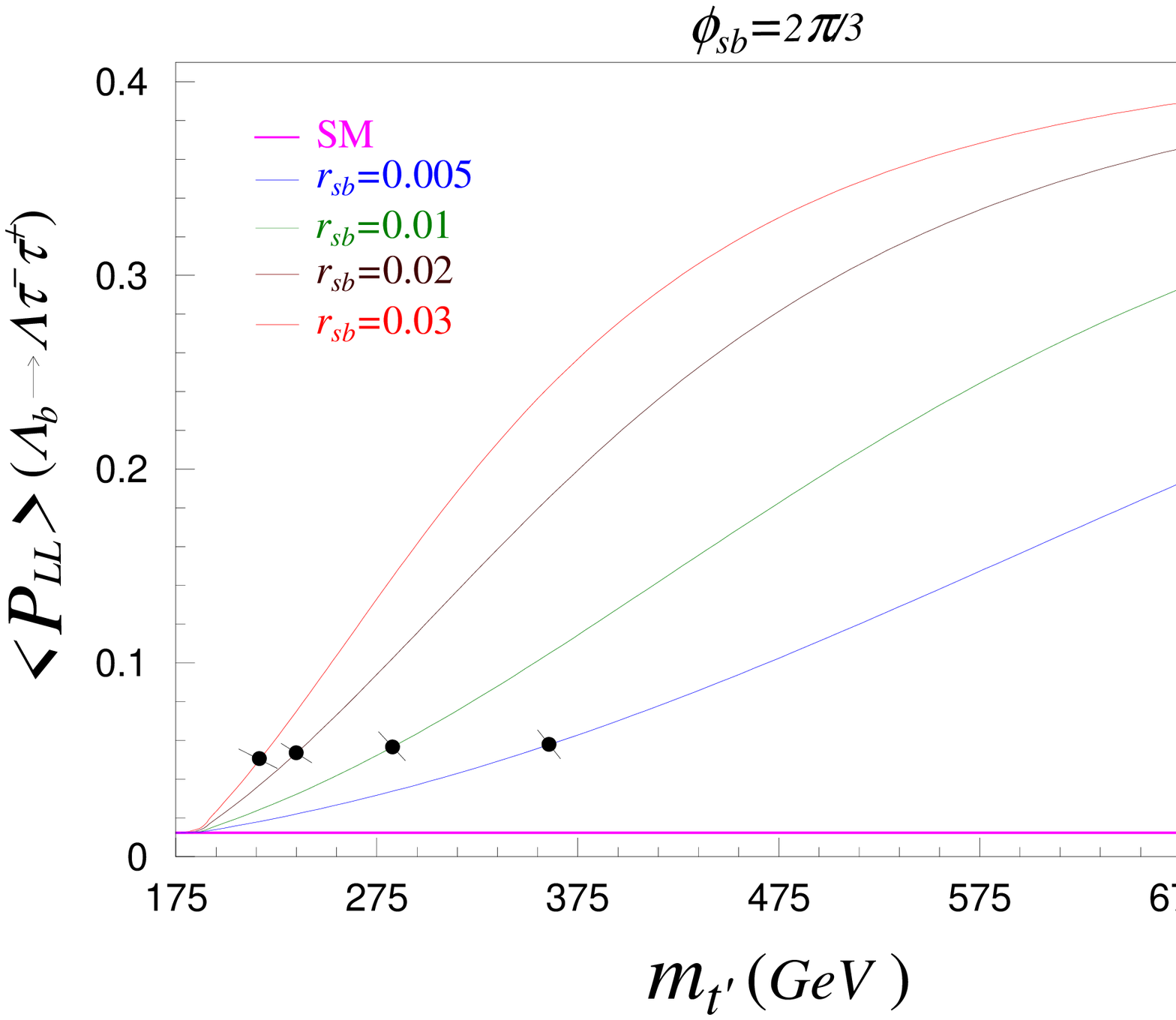}
\vskip 7.8cm \caption{}
\end{figure}

\begin{figure}
\vskip 2.5 cm
    \includegraphics{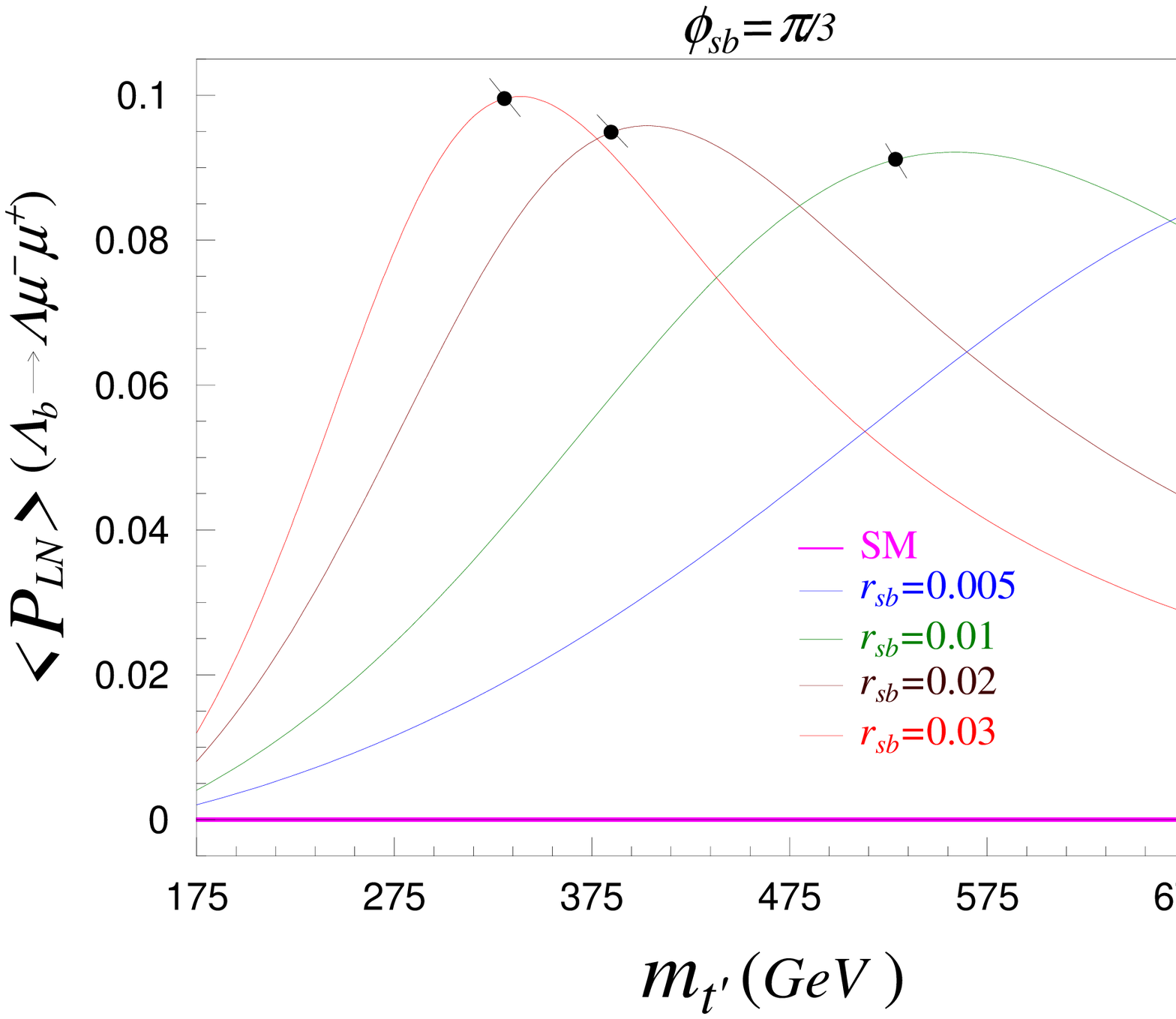}
\vskip 7.8 cm \caption{}
\end{figure}

\begin{figure}
\vskip 2.5 cm
    \includegraphics{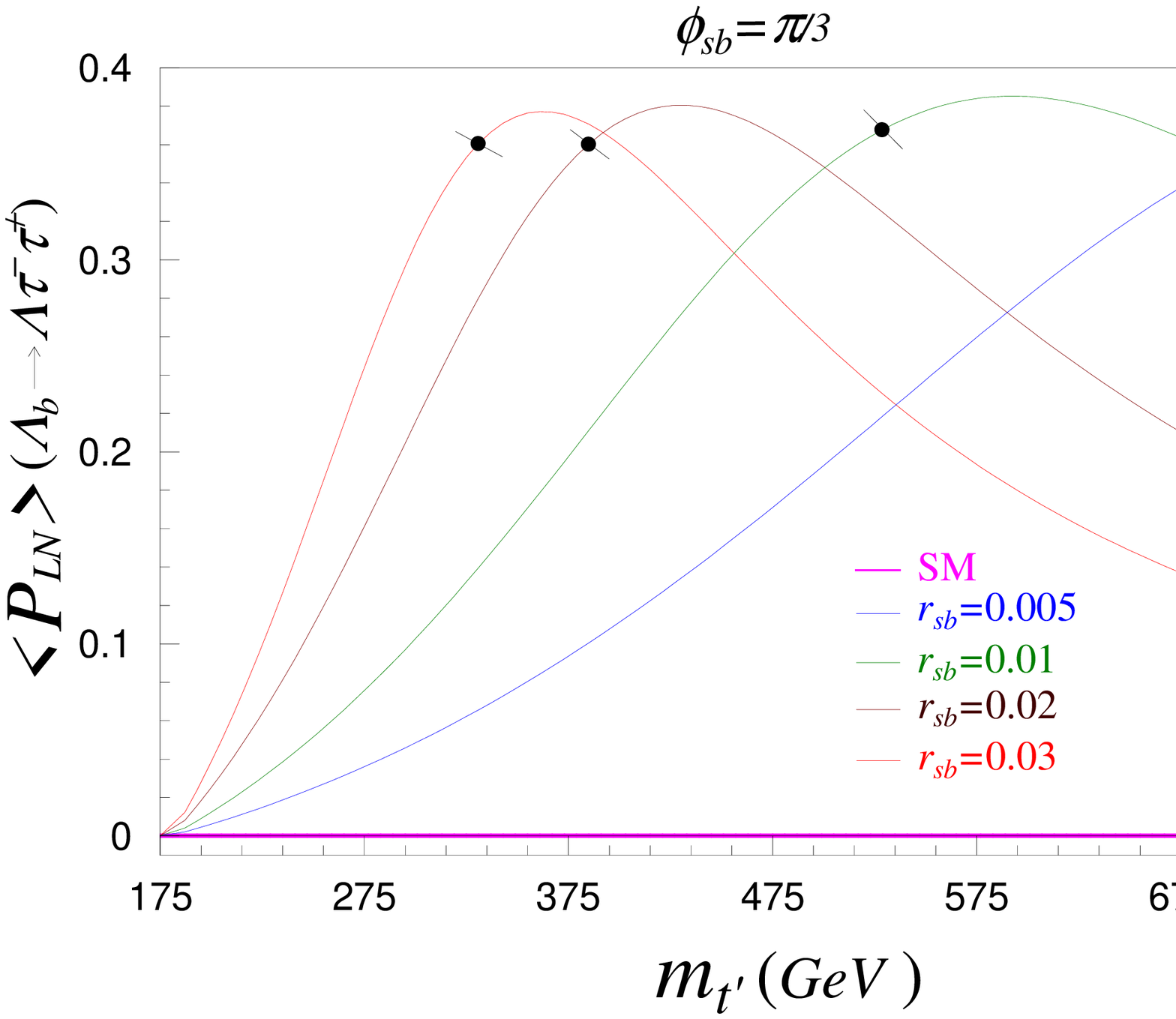}
\vskip 7.8 cm \caption{}
\end{figure}

\begin{figure}
\vskip 2.5 cm
    \includegraphics{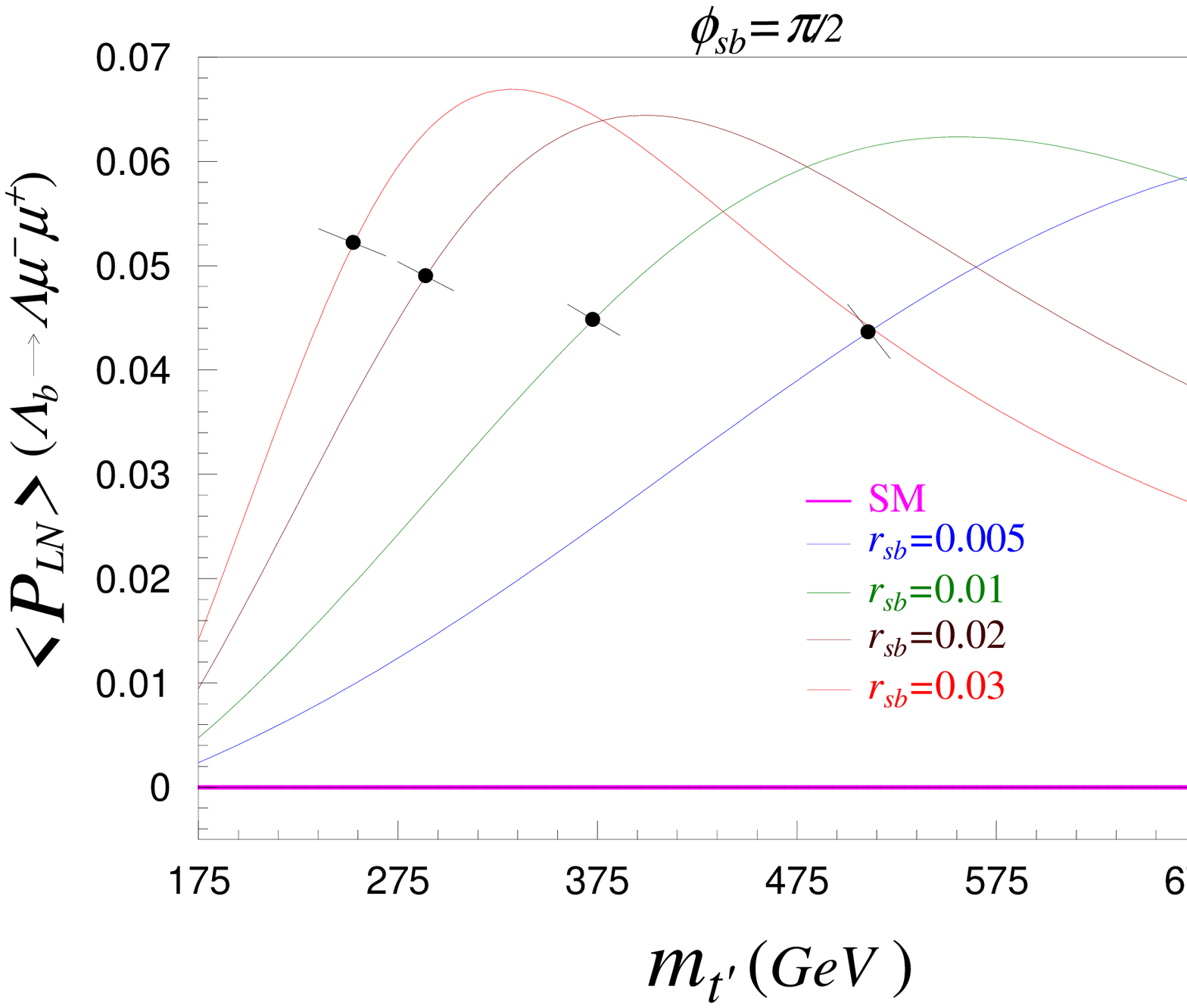}
\vskip 7.8 cm
\caption{}
\end{figure}

\begin{figure}
\vskip 2.5 cm
    \includegraphics{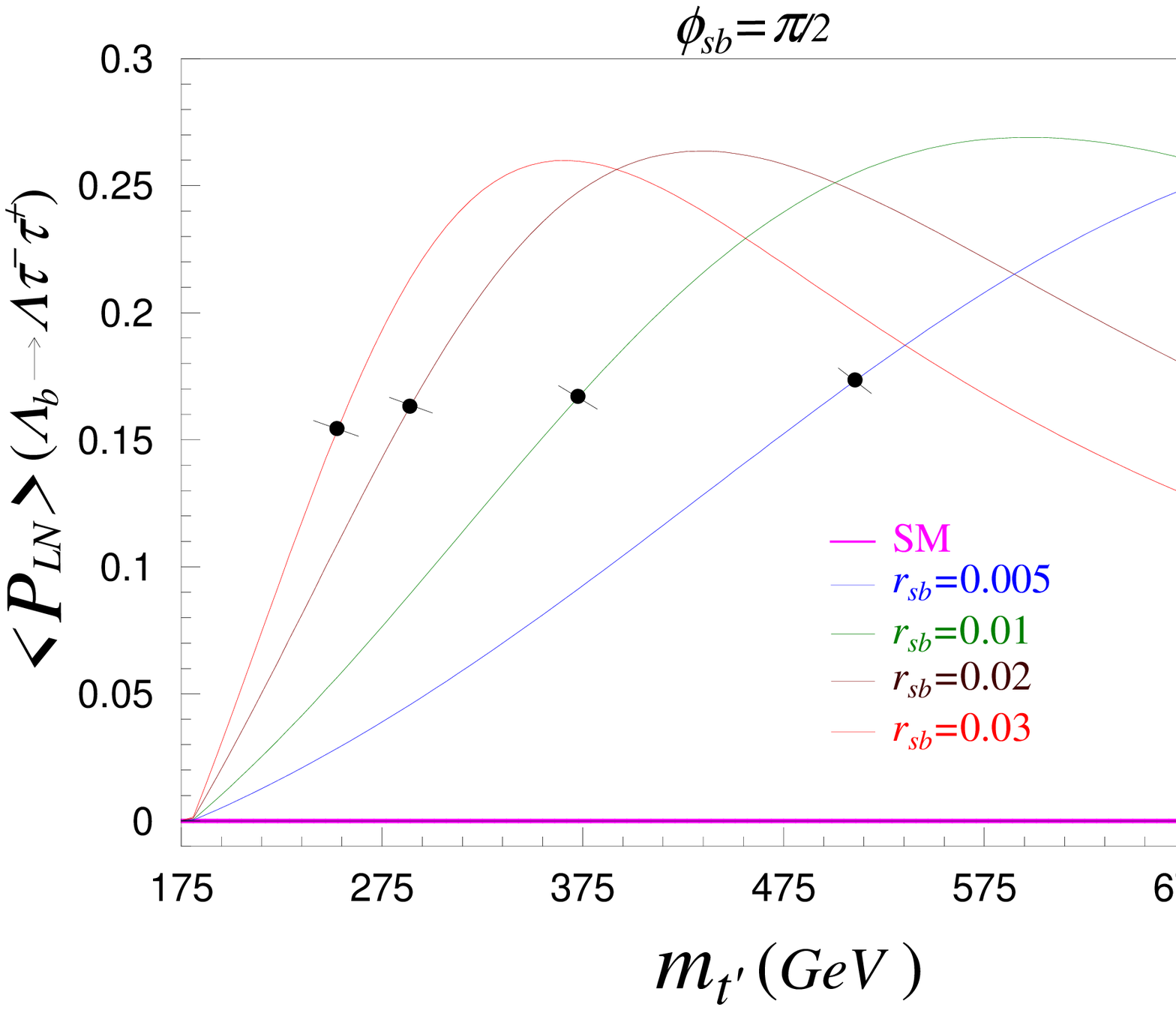}
\vskip 7.8 cm \caption{}
\end{figure}

\begin{figure}
\vskip 2.5 cm
    \includegraphics{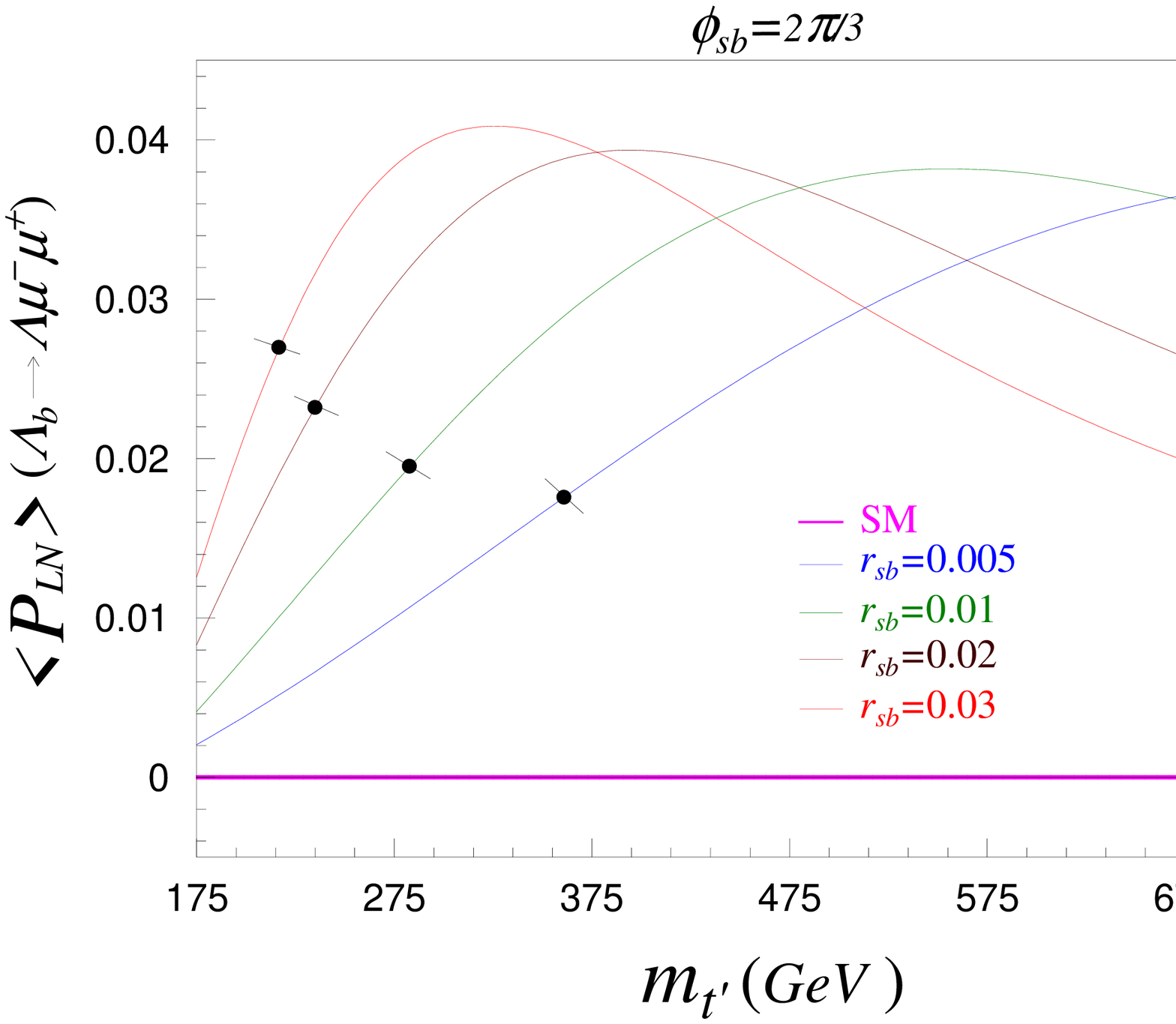}
\vskip 7.8 cm \caption{}
\end{figure}

\begin{figure}
\vskip 2.5 cm
    \includegraphics{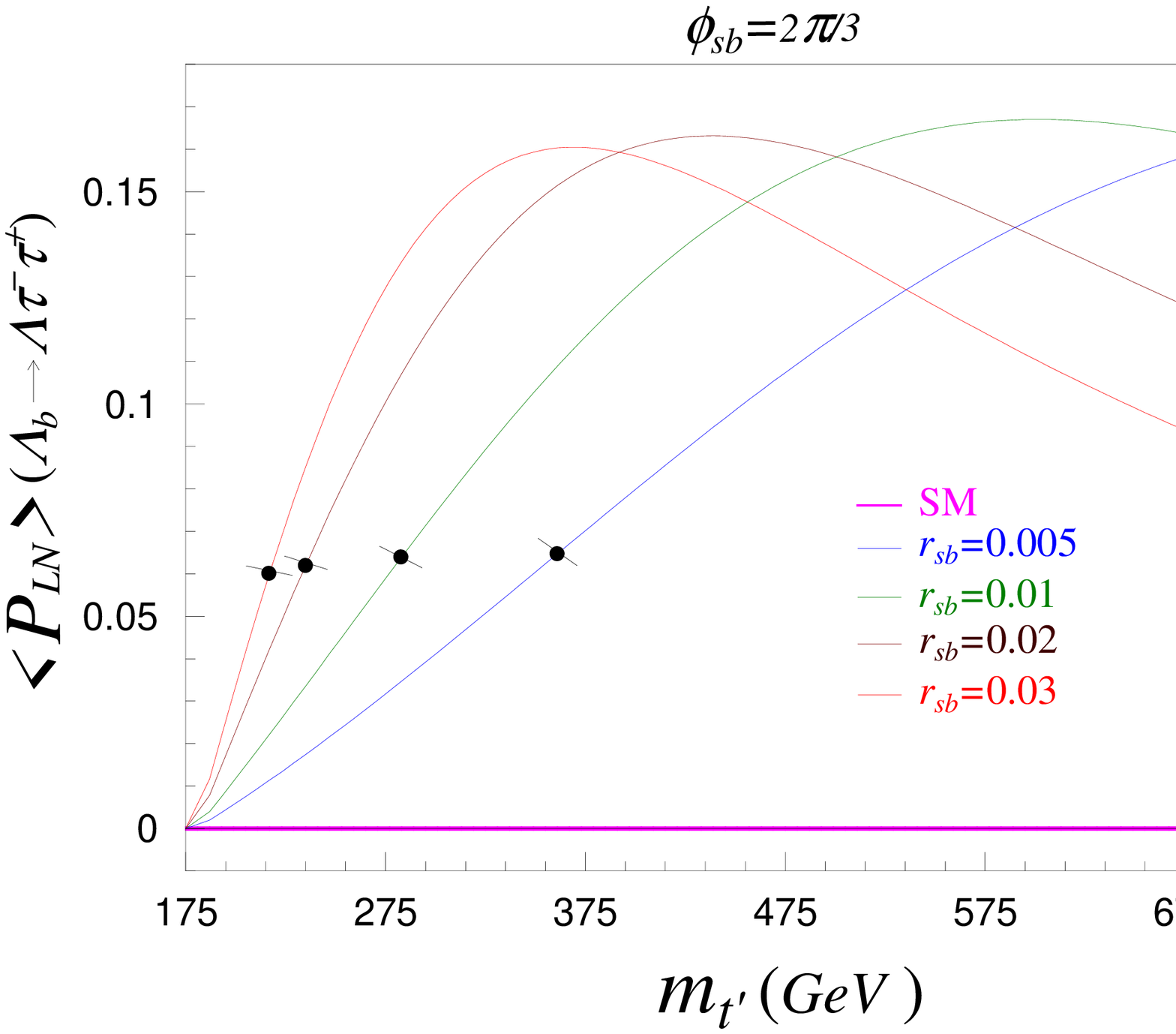}
\vskip 7.8 cm \caption{}
\end{figure}

\begin{figure}
\vskip 2.5 cm
    \includegraphics{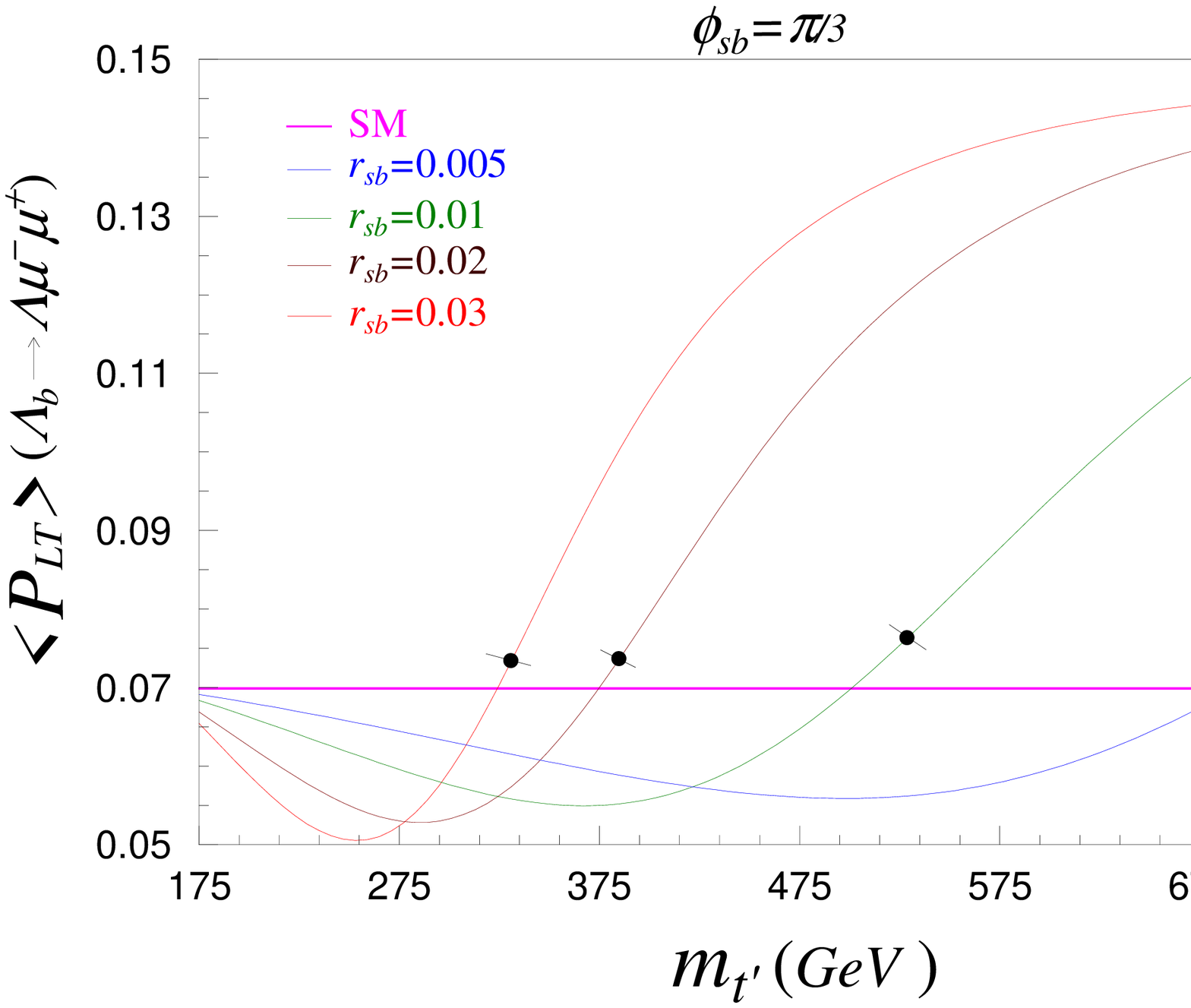}
\vskip 7.8cm \caption{}
\end{figure}

\begin{figure}
\vskip 2.5 cm
    \includegraphics{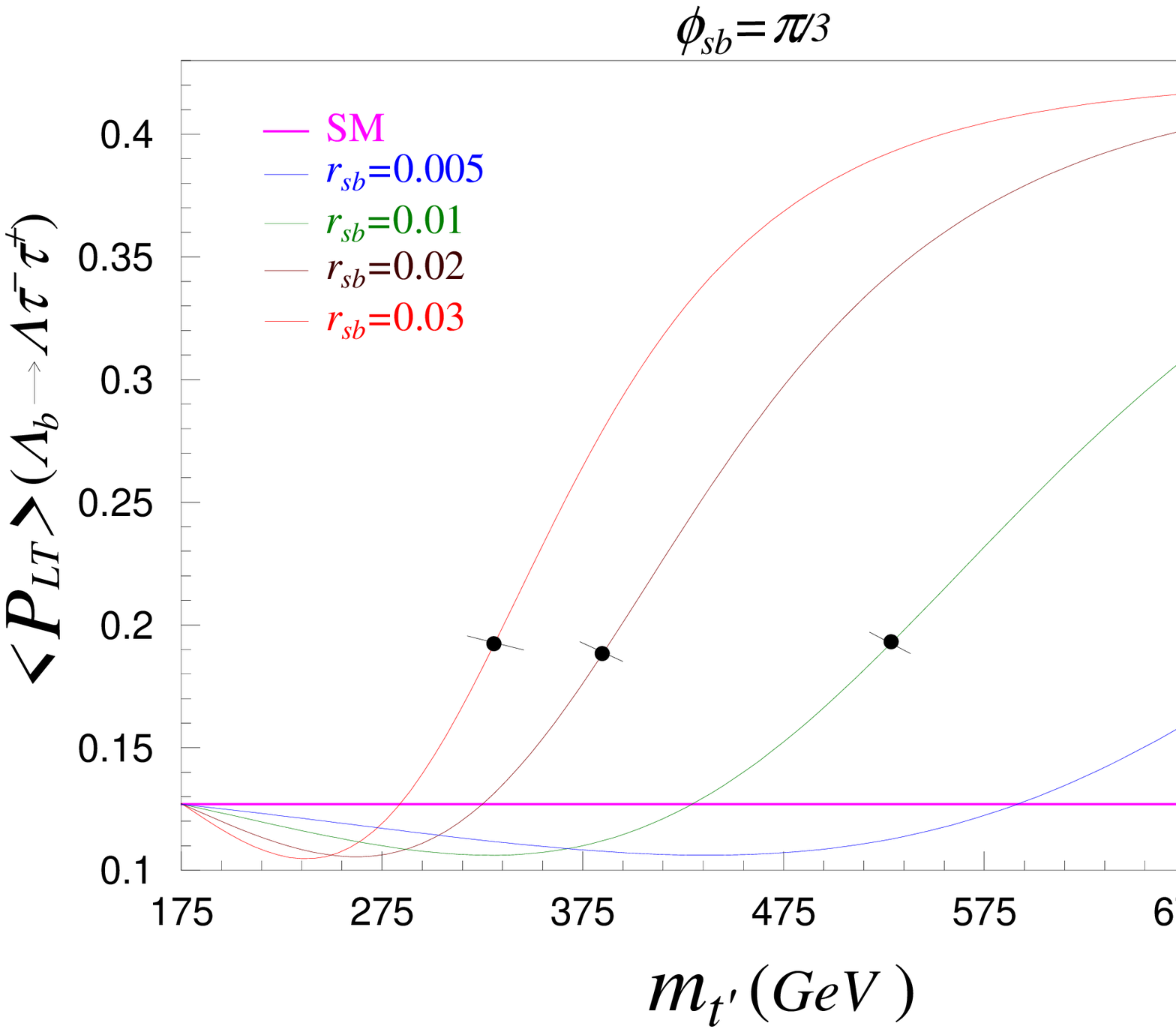}
\vskip 7.8cm \caption{}
\end{figure}

\begin{figure}
\vskip 2.5 cm
    \includegraphics{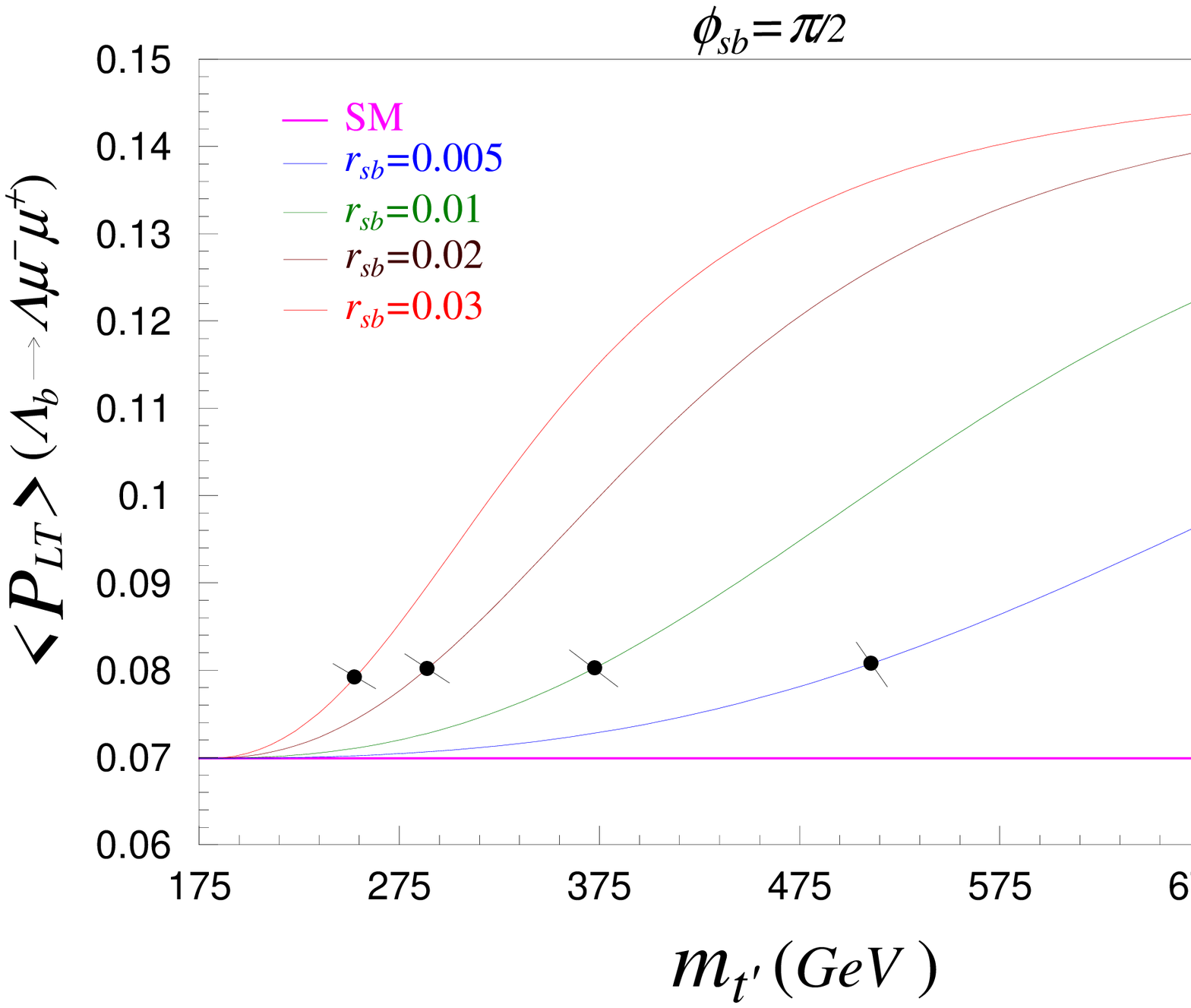}
\vskip 7.8cm \caption{}
\end{figure}

\begin{figure}
\vskip 2.5 cm
    \includegraphics{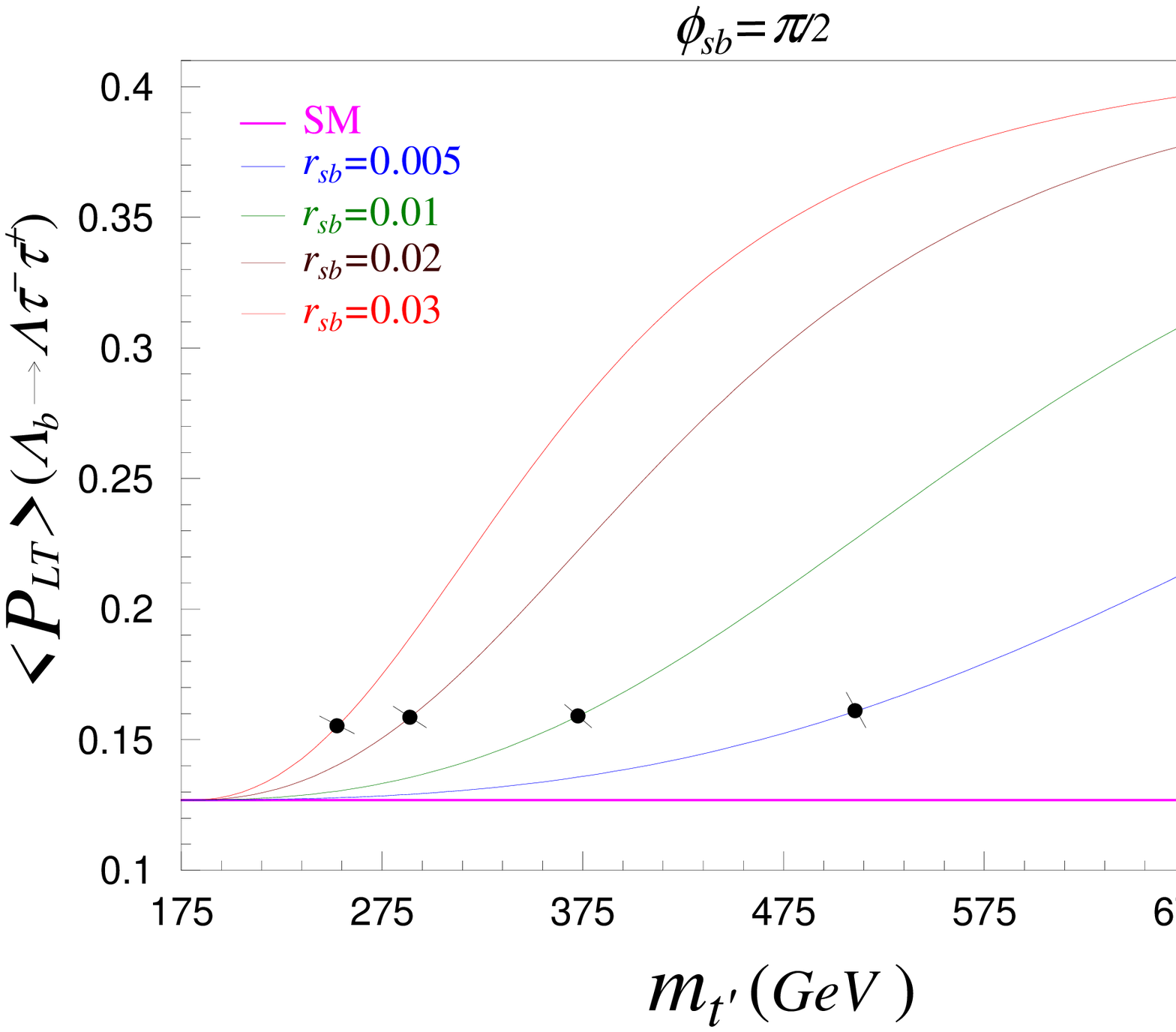}
\vskip 7.8cm \caption{}
\end{figure}

\begin{figure}
\vskip 2.5 cm
    \includegraphics{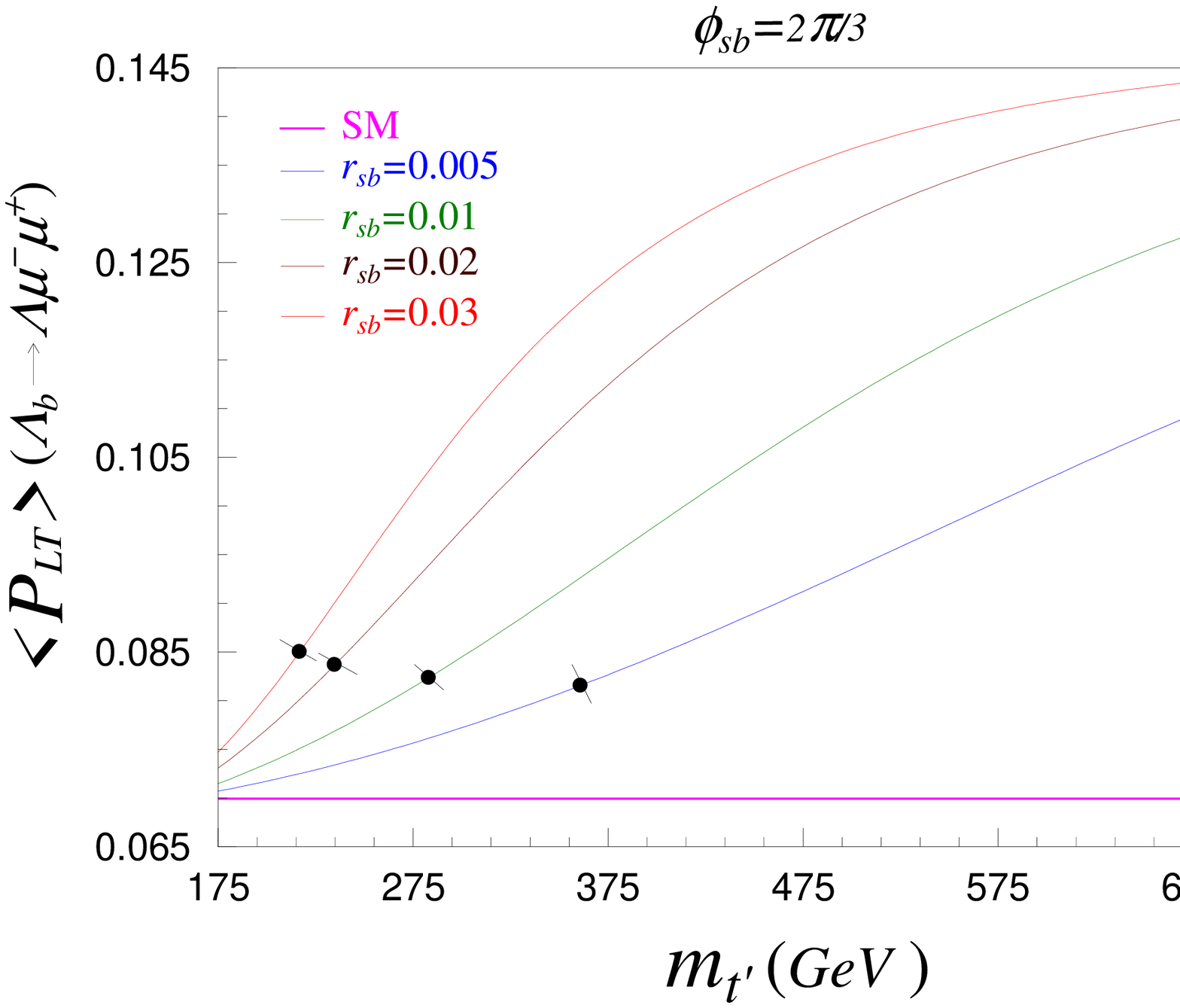}
\vskip 7.8cm \caption{}
\end{figure}

\begin{figure}
\vskip 2.5 cm
    \includegraphics{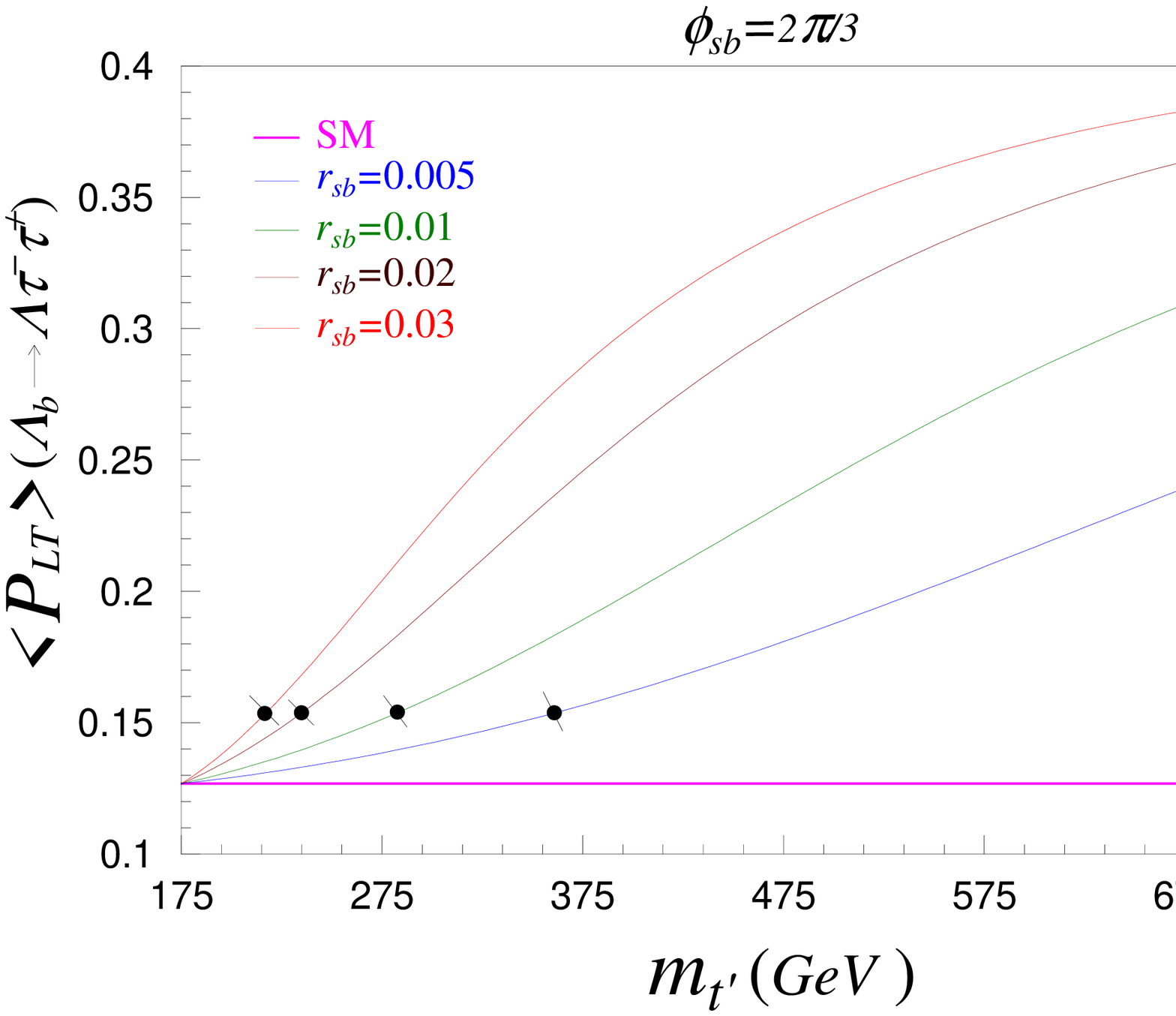}
\vskip 7.8cm \caption{}
\end{figure}

\begin{figure}
\vskip 2.5 cm
    \includegraphics{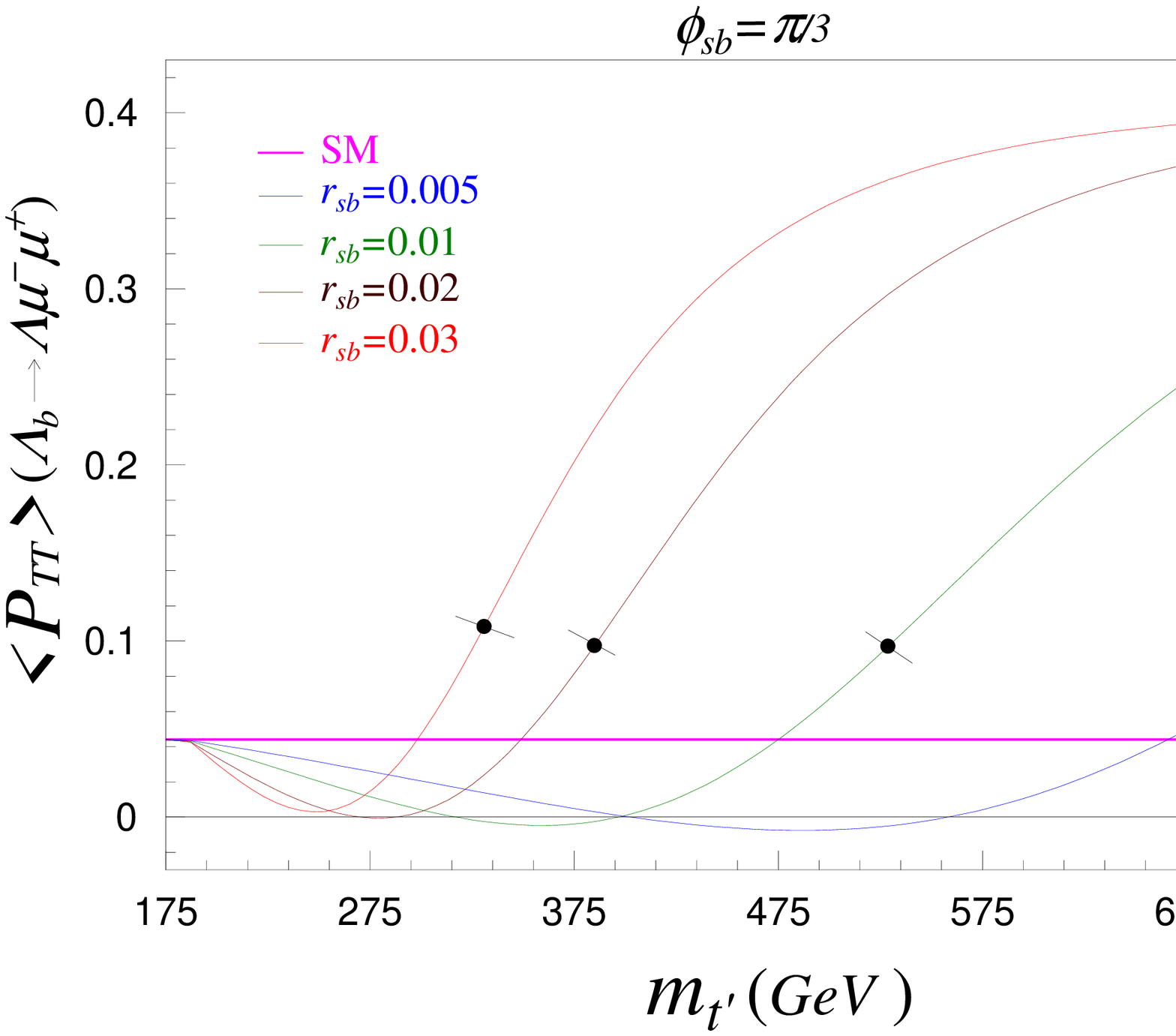}
\vskip 7.8cm \caption{}
\end{figure}

\begin{figure}
\vskip 2.5 cm
    \includegraphics{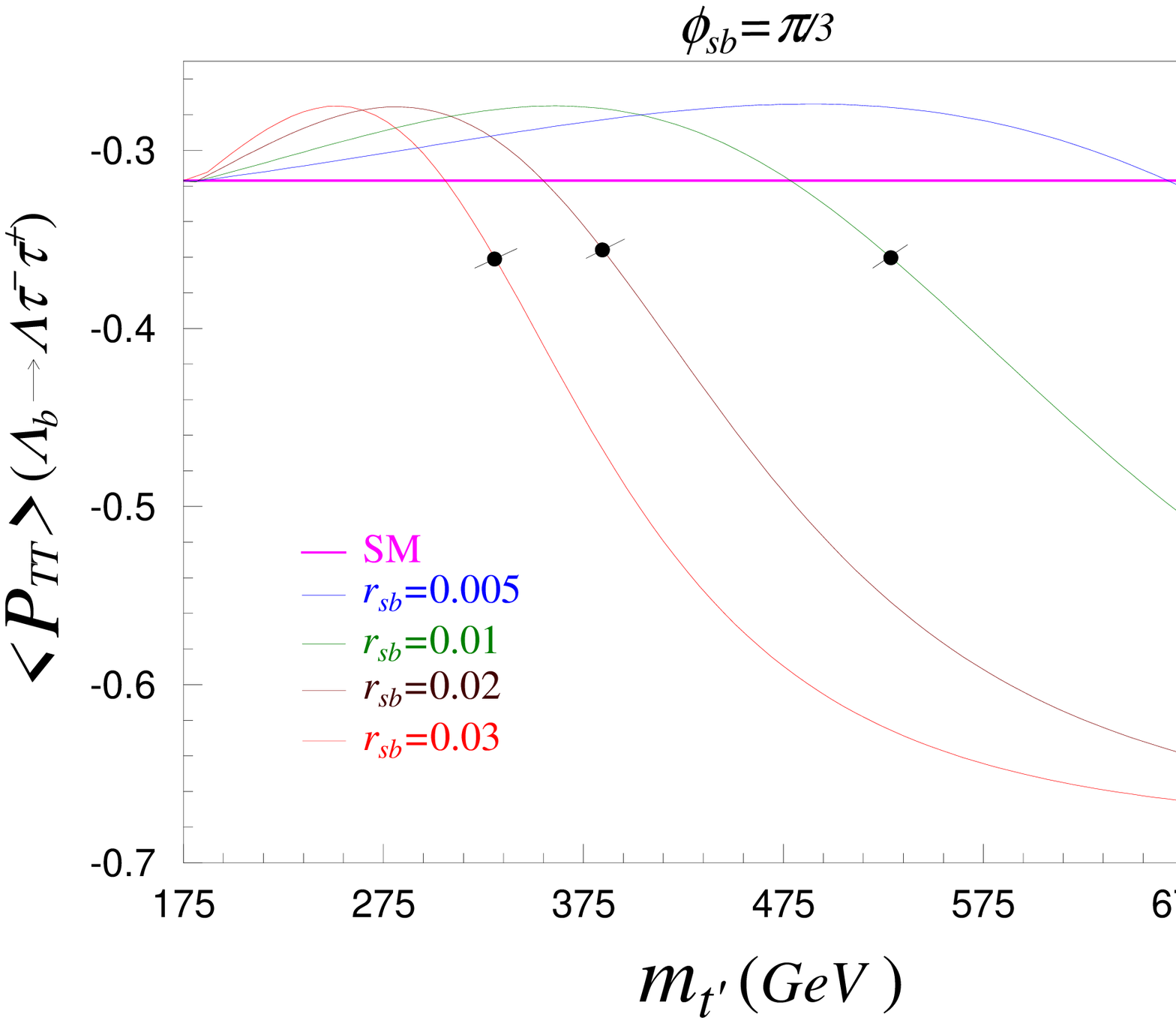}
\vskip 7.8cm \caption{}
\end{figure}

\begin{figure}
\vskip 2.5 cm
    \includegraphics{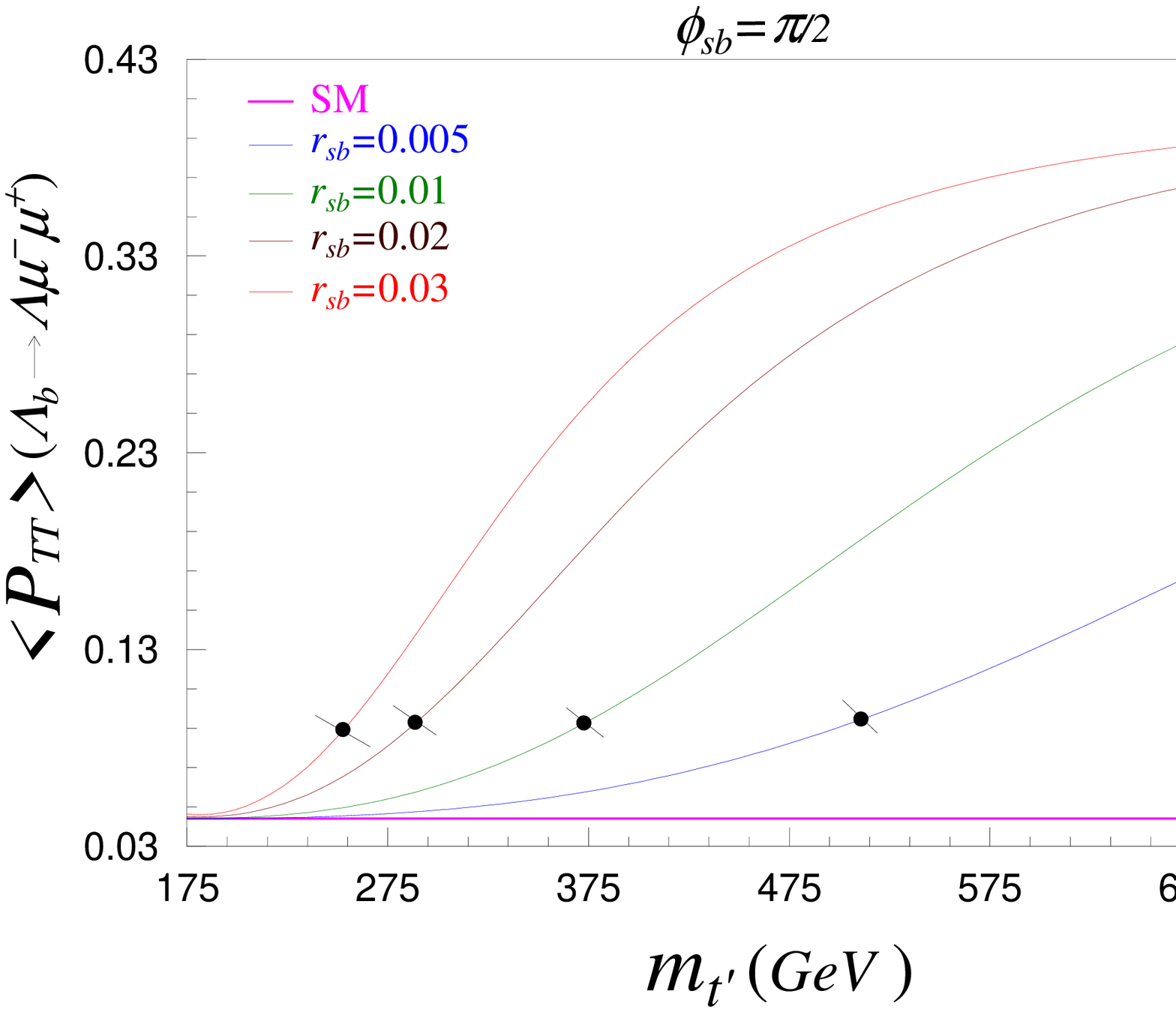}
\vskip 7.8cm \caption{}
\end{figure}

\begin{figure}
\vskip 2.5 cm
    \includegraphics{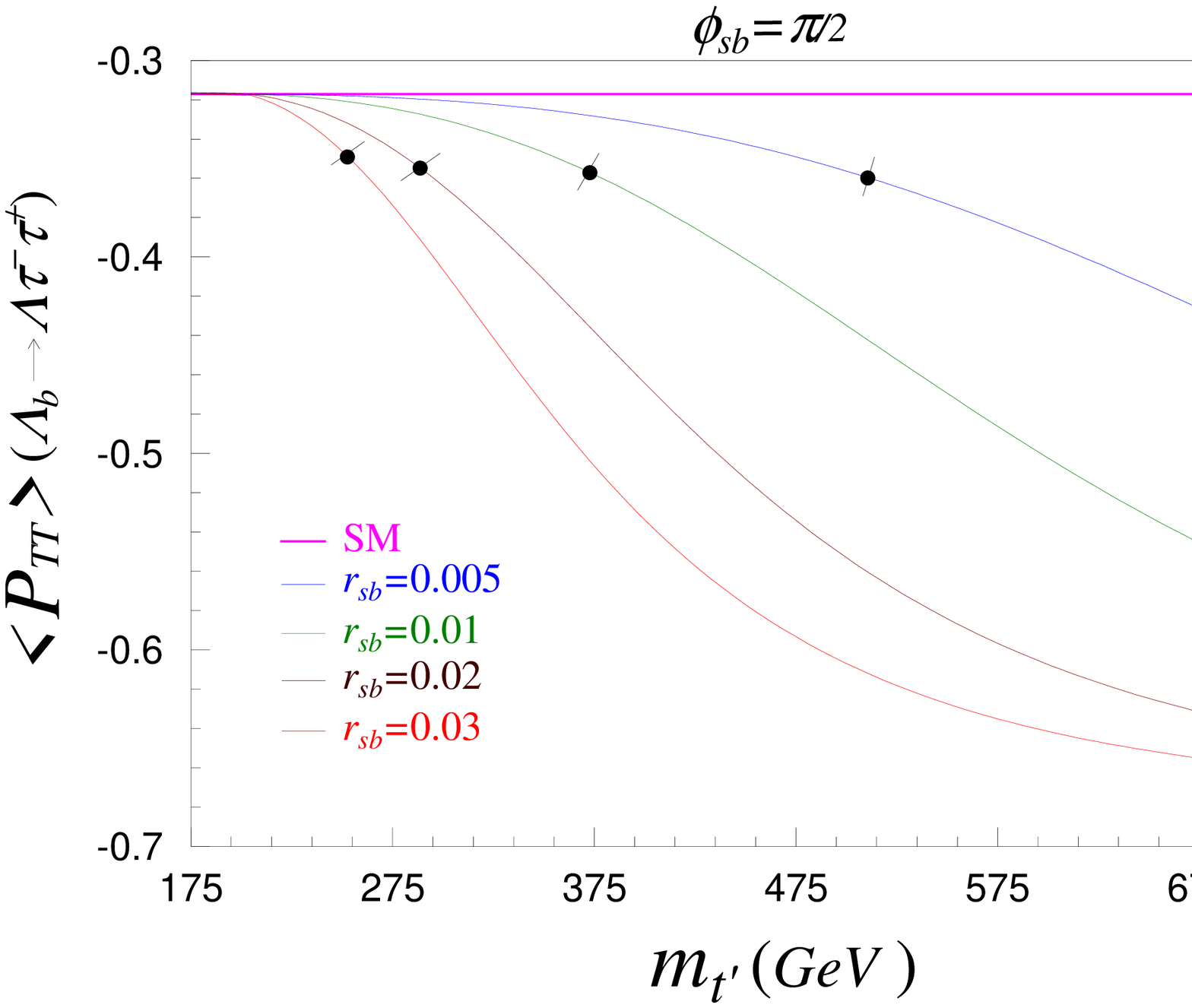}
\vskip 7.8cm \caption{}
\end{figure}

\begin{figure}
\vskip 2.5 cm
    \includegraphics{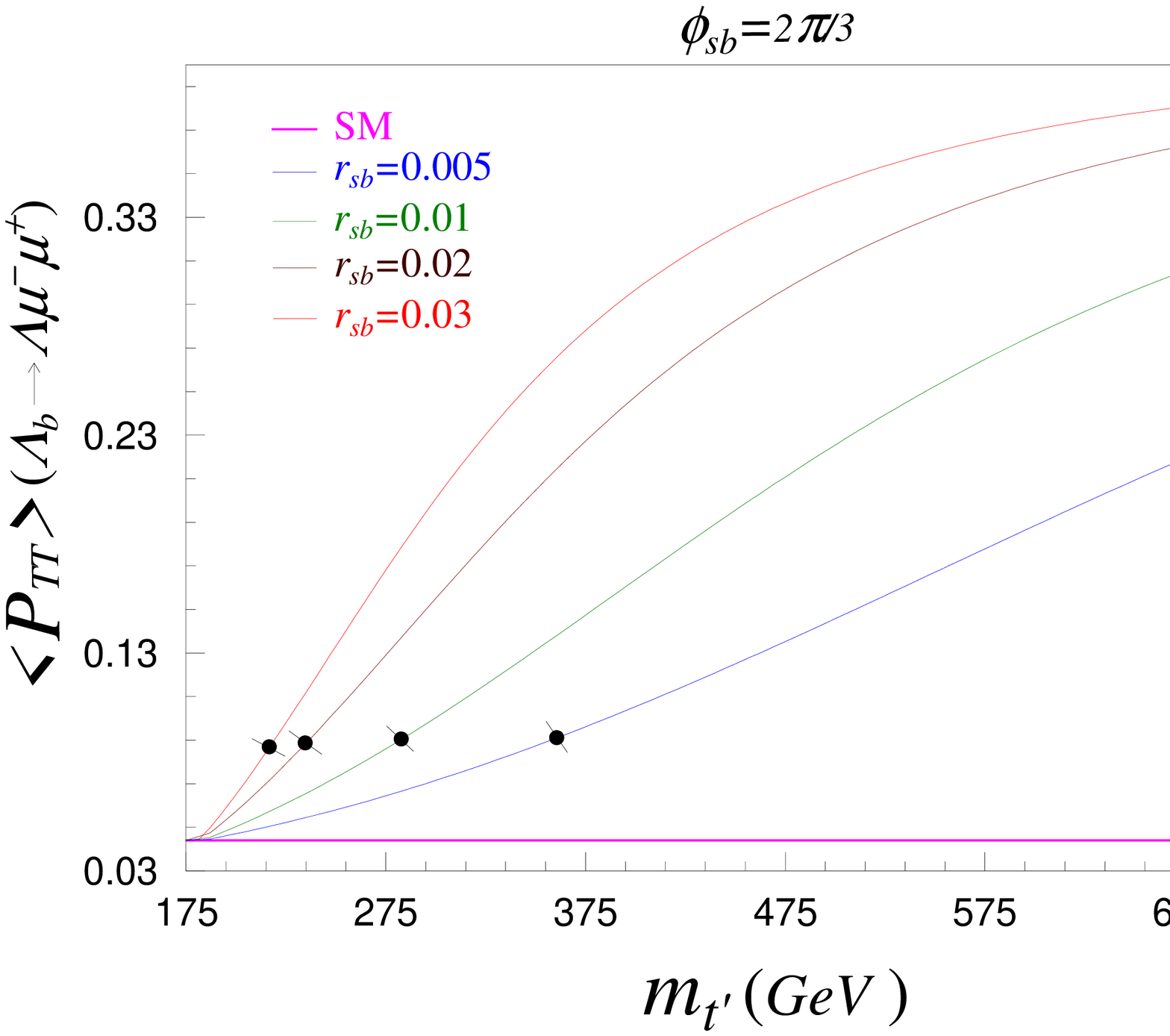}
\vskip 7.8cm \caption{}
\end{figure}

\begin{figure}
\vskip 2.5 cm
    \includegraphics{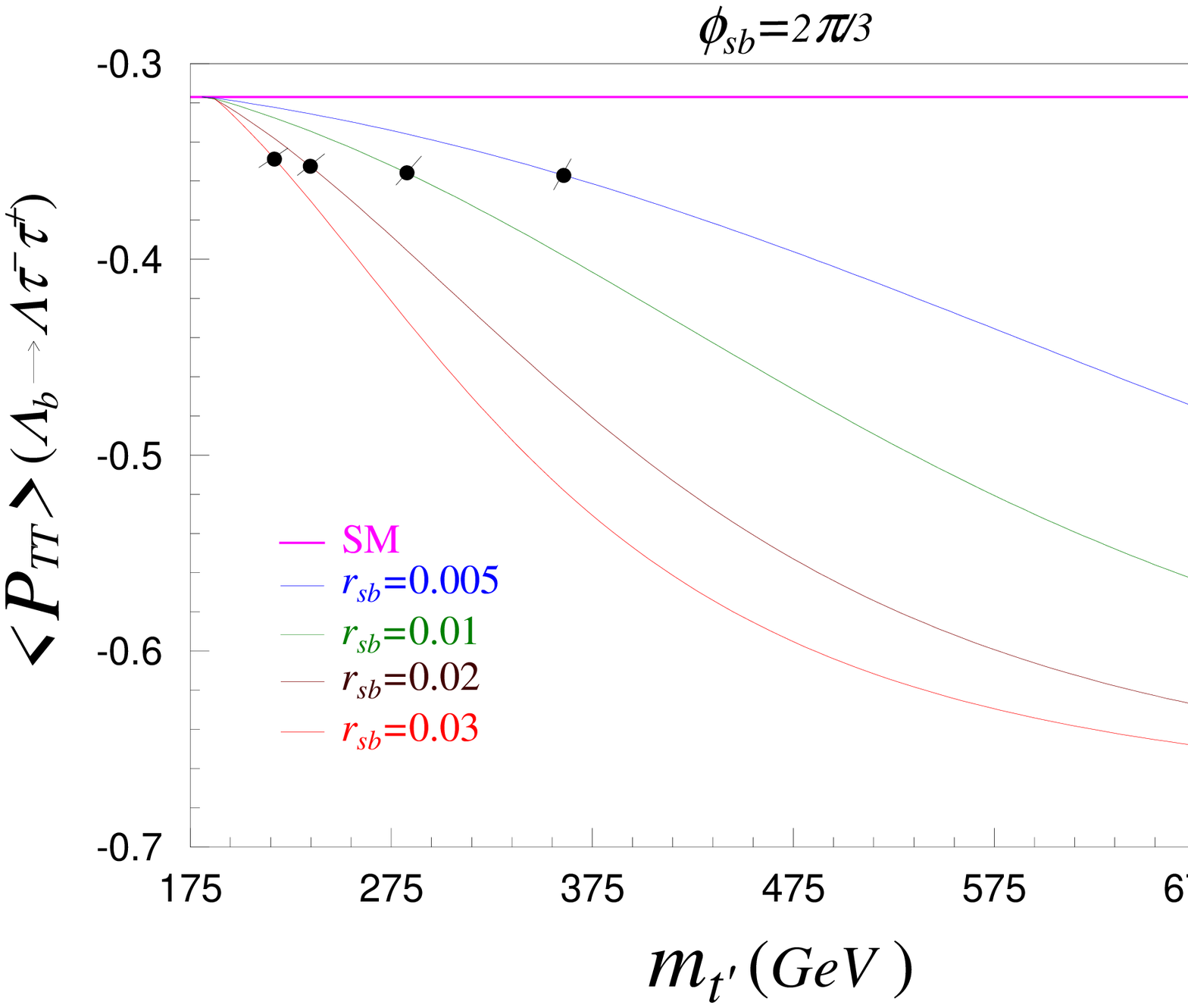}
\vskip 7.8cm \caption{}
\end{figure}

\begin{figure}
\vskip 2.5 cm
    \includegraphics{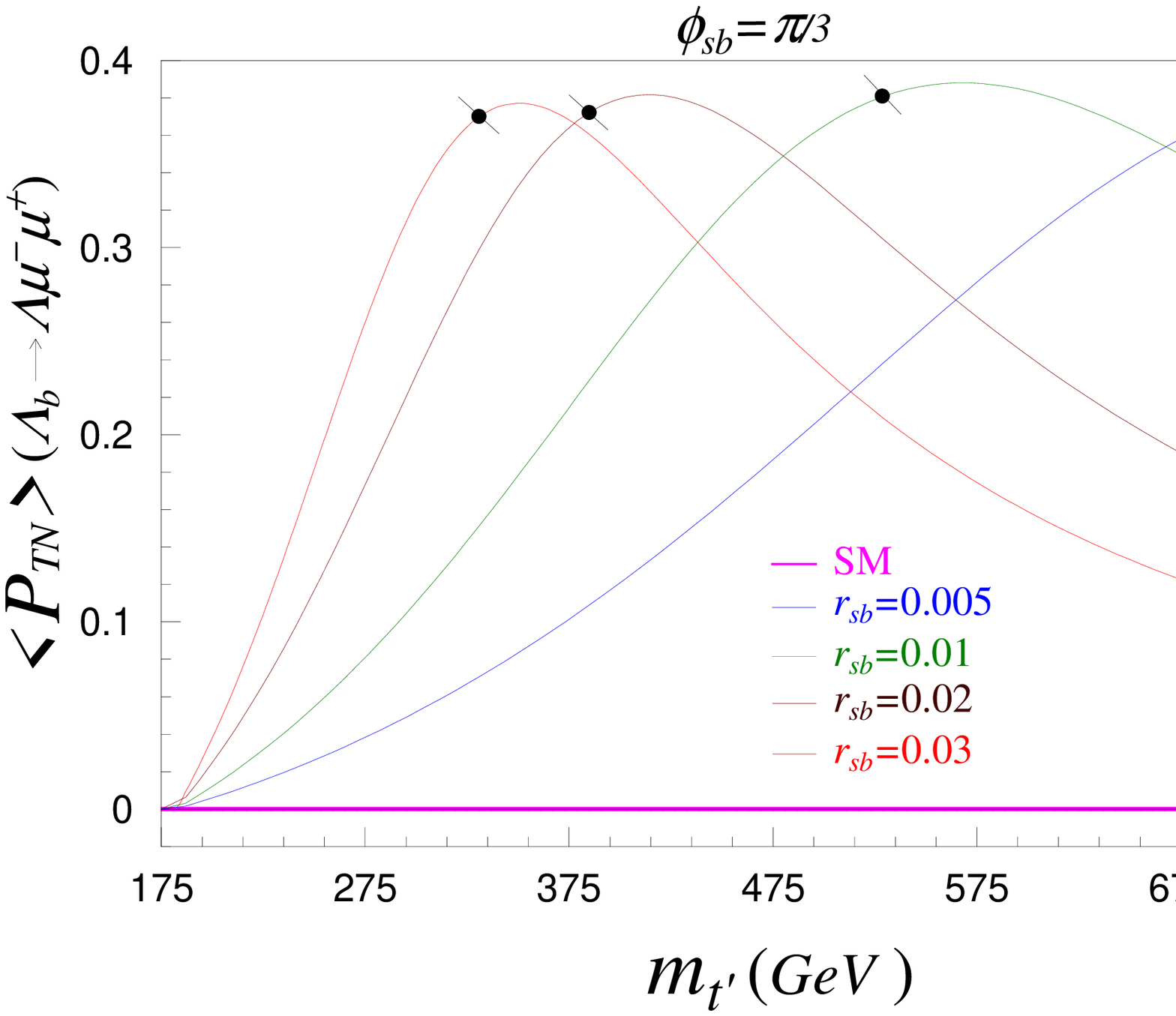}
\vskip 7.8 cm \caption{}
\end{figure}

\begin{figure}
\vskip 2.5 cm
    \includegraphics{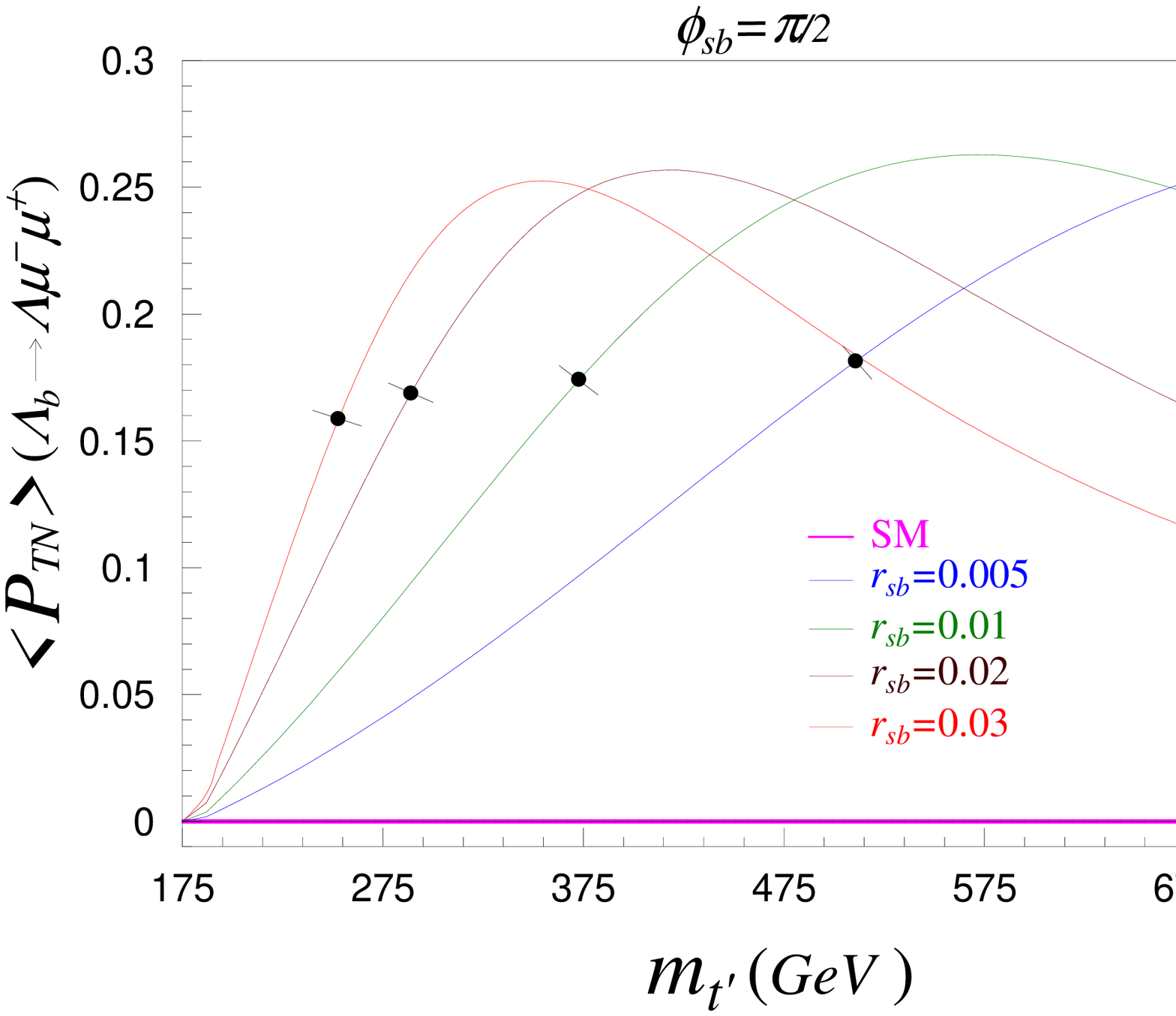}
\vskip 7.8 cm \caption{}
\end{figure}

\begin{figure}
\vskip 2.5 cm
    \includegraphics{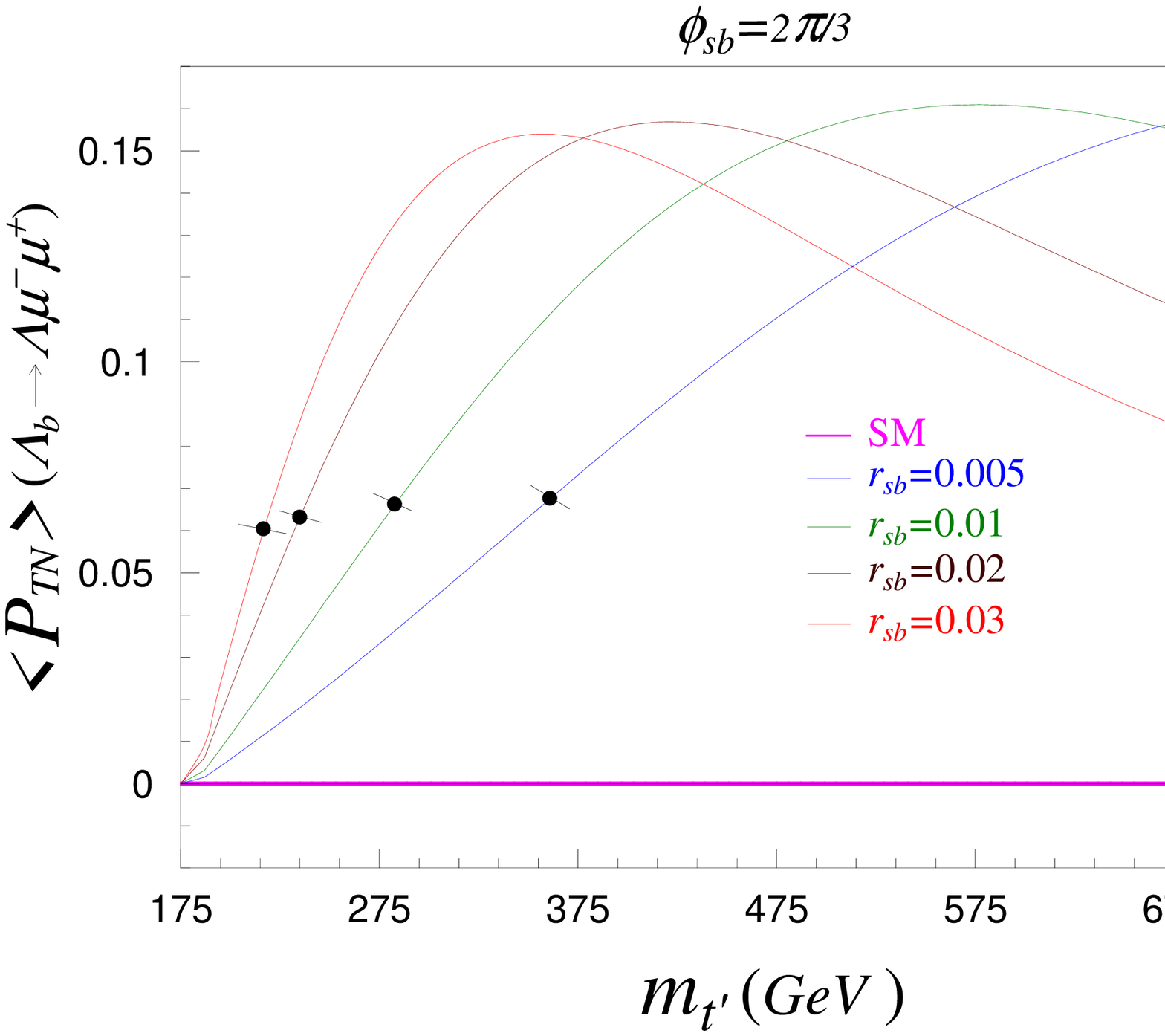}
\vskip 7.8 cm \caption{}
\end{figure}

\begin{figure}
\vskip 2.5 cm
    \includegraphics{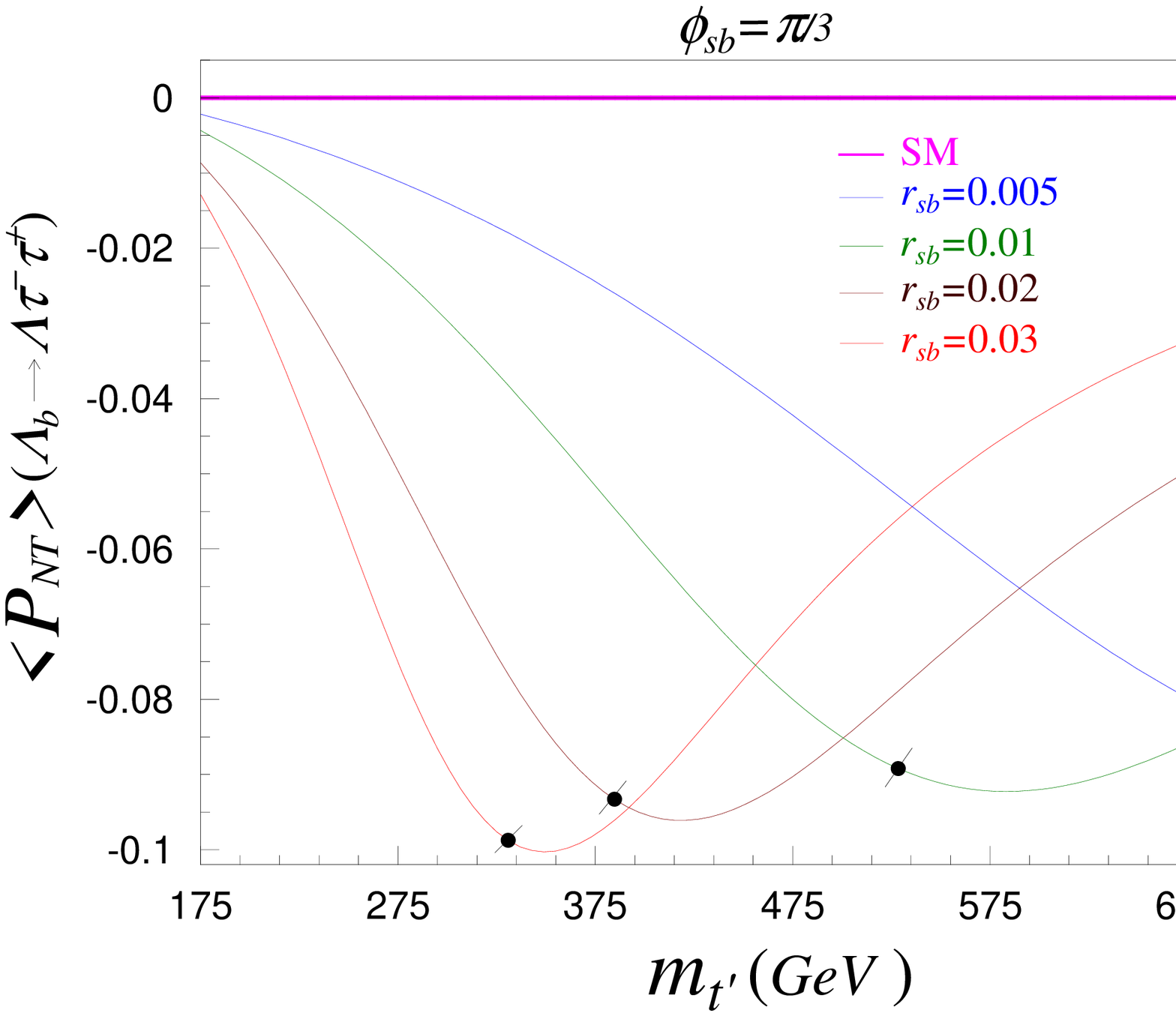}
\vskip 7.8 cm \caption{}
\end{figure}

\begin{figure}
\vskip 2.5 cm
    \includegraphics{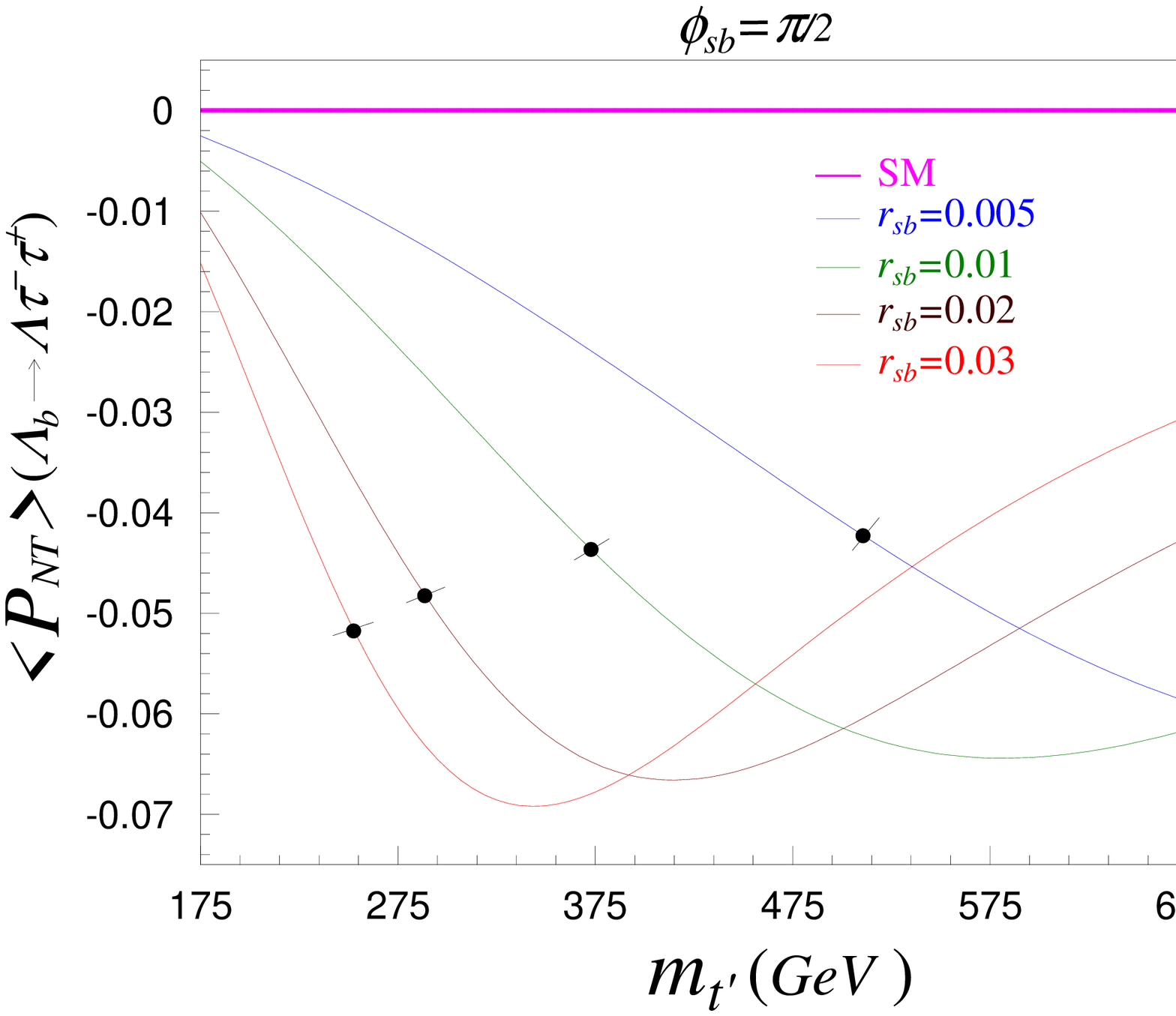}
\vskip 7.8 cm \caption{}
\end{figure}

\begin{figure}
\vskip 2.5 cm
    \includegraphics{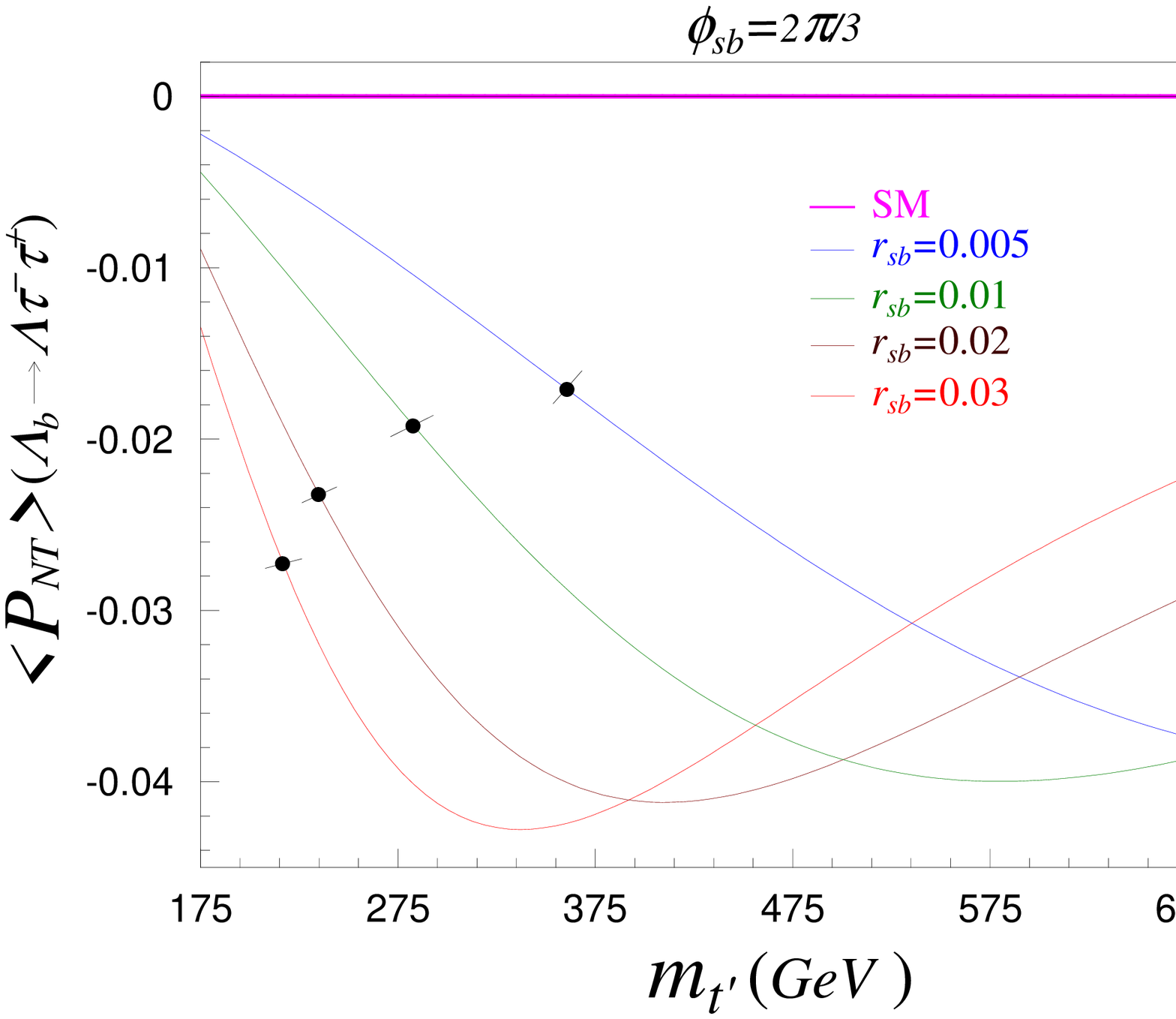}
\vskip 7.8 cm \caption{}
\end{figure}

\begin{figure}
\vskip 2.5 cm
    \includegraphics{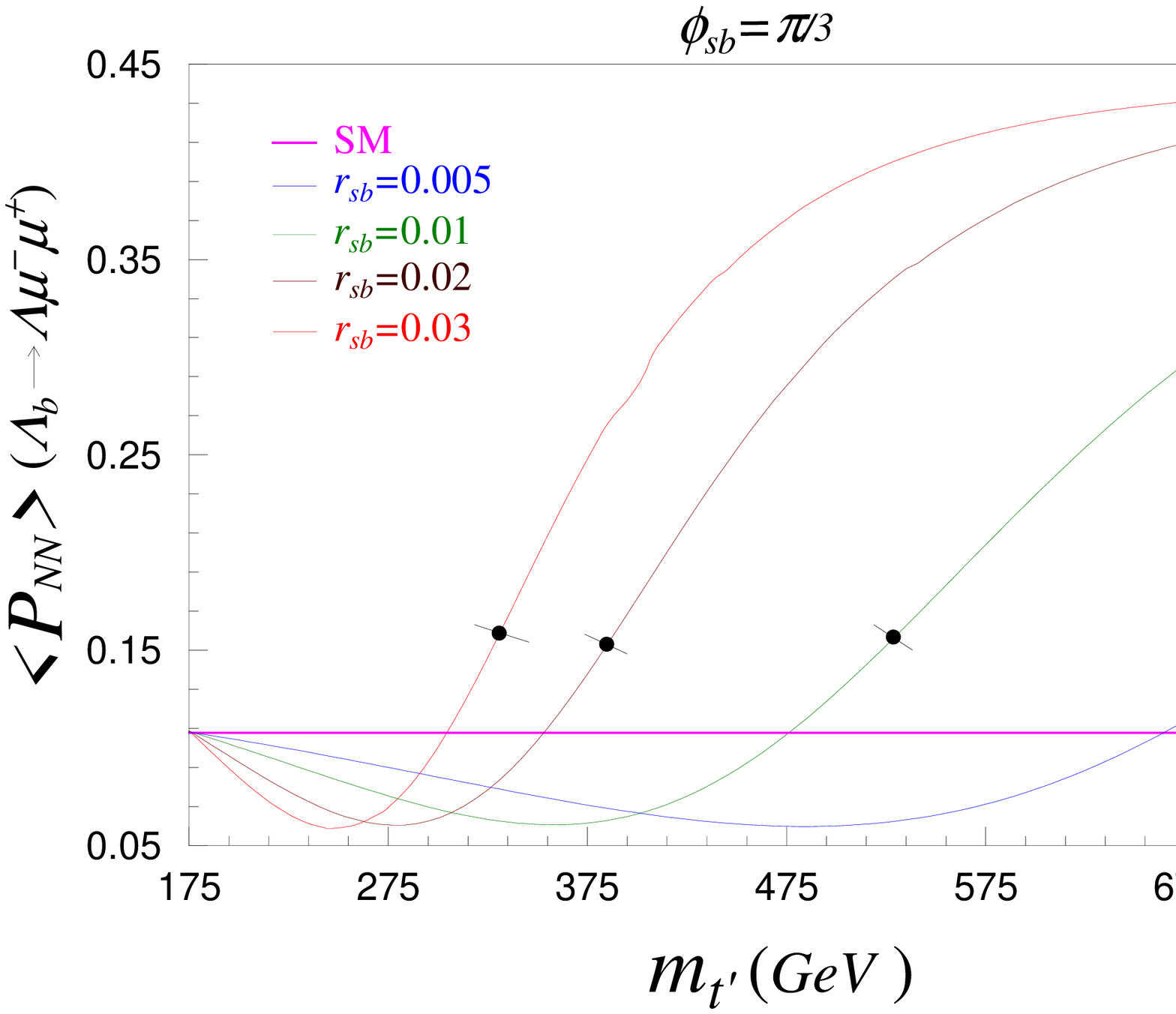}
\vskip 7.8cm \caption{}
\end{figure}

\begin{figure}
\vskip 2.5 cm
    \includegraphics{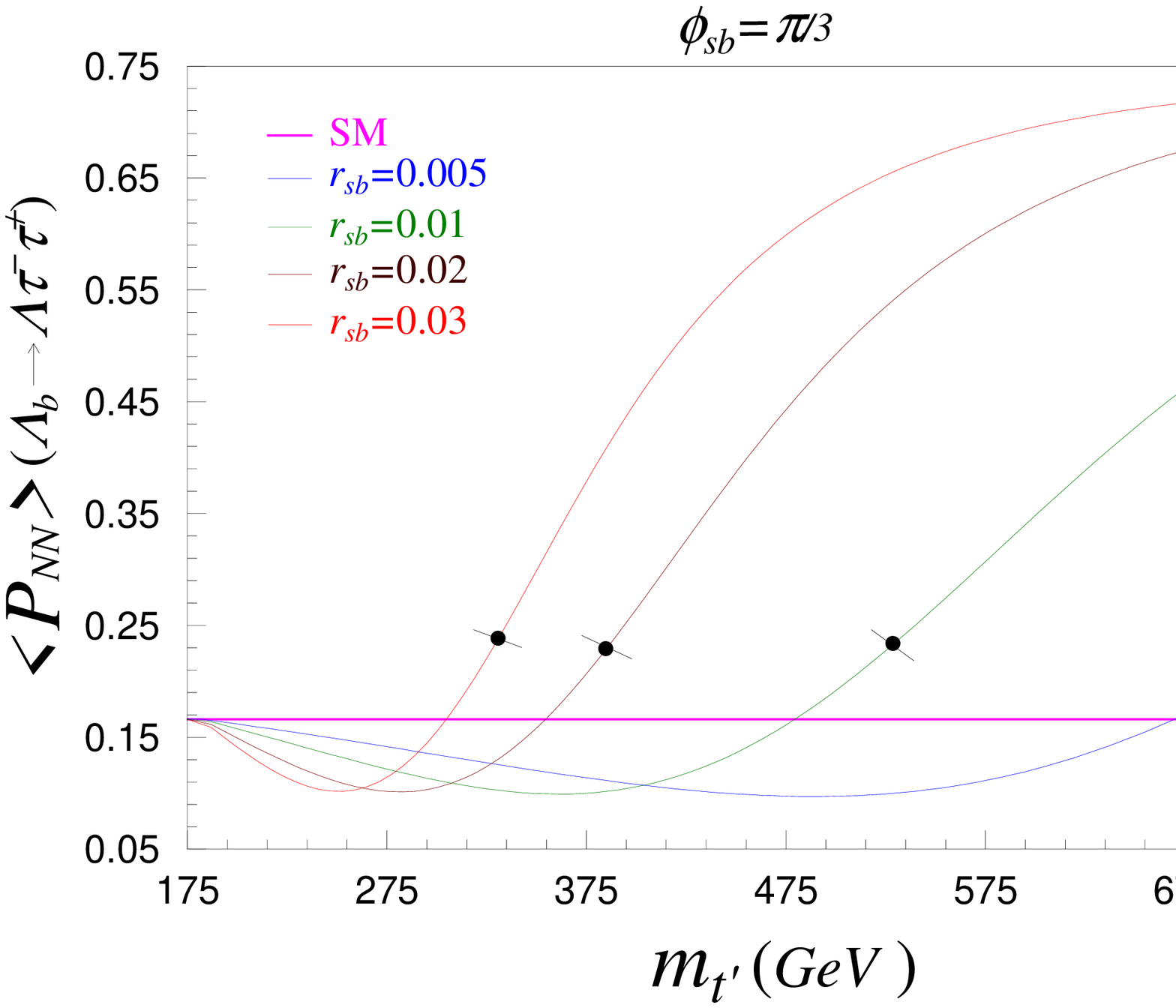}
\vskip 7.8cm \caption{}
\end{figure}

\begin{figure}
\vskip 2.5 cm
    \includegraphics{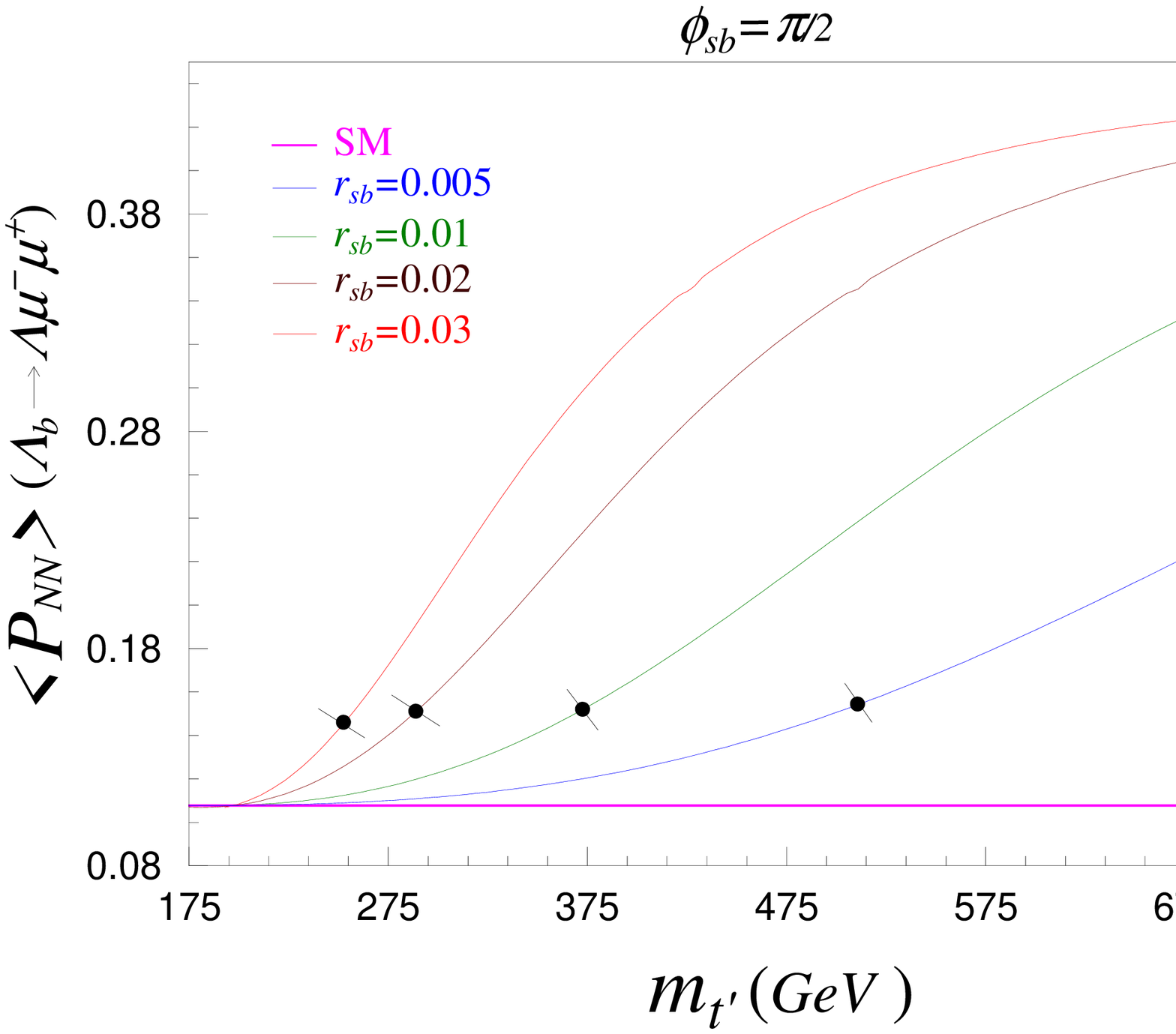}
\vskip 7.8cm \caption{}
\end{figure}

\begin{figure}
\vskip 2.5 cm
    \includegraphics{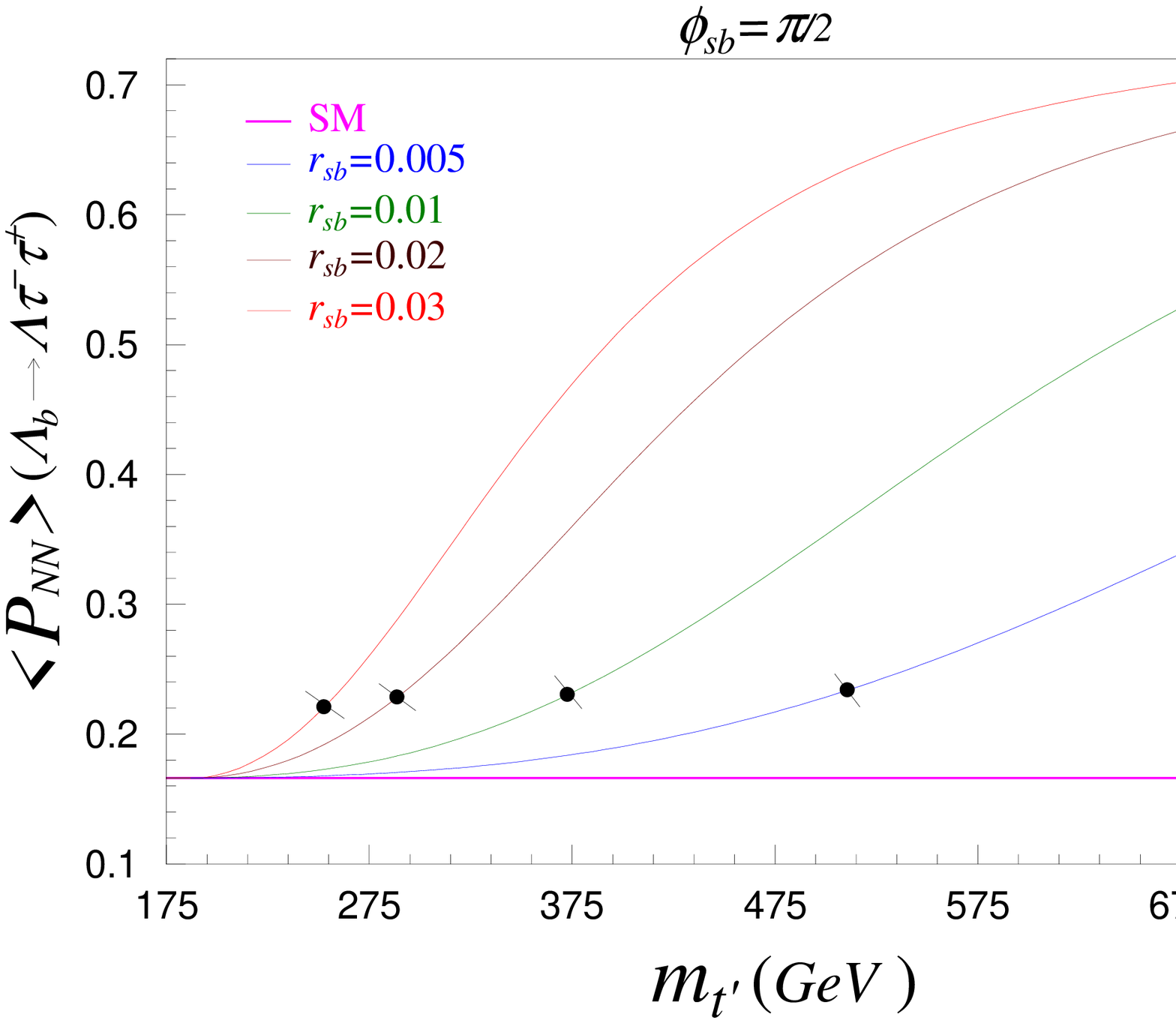}
\vskip 7.8cm \caption{}
\end{figure}

\begin{figure}
\vskip 2.5 cm
    \includegraphics{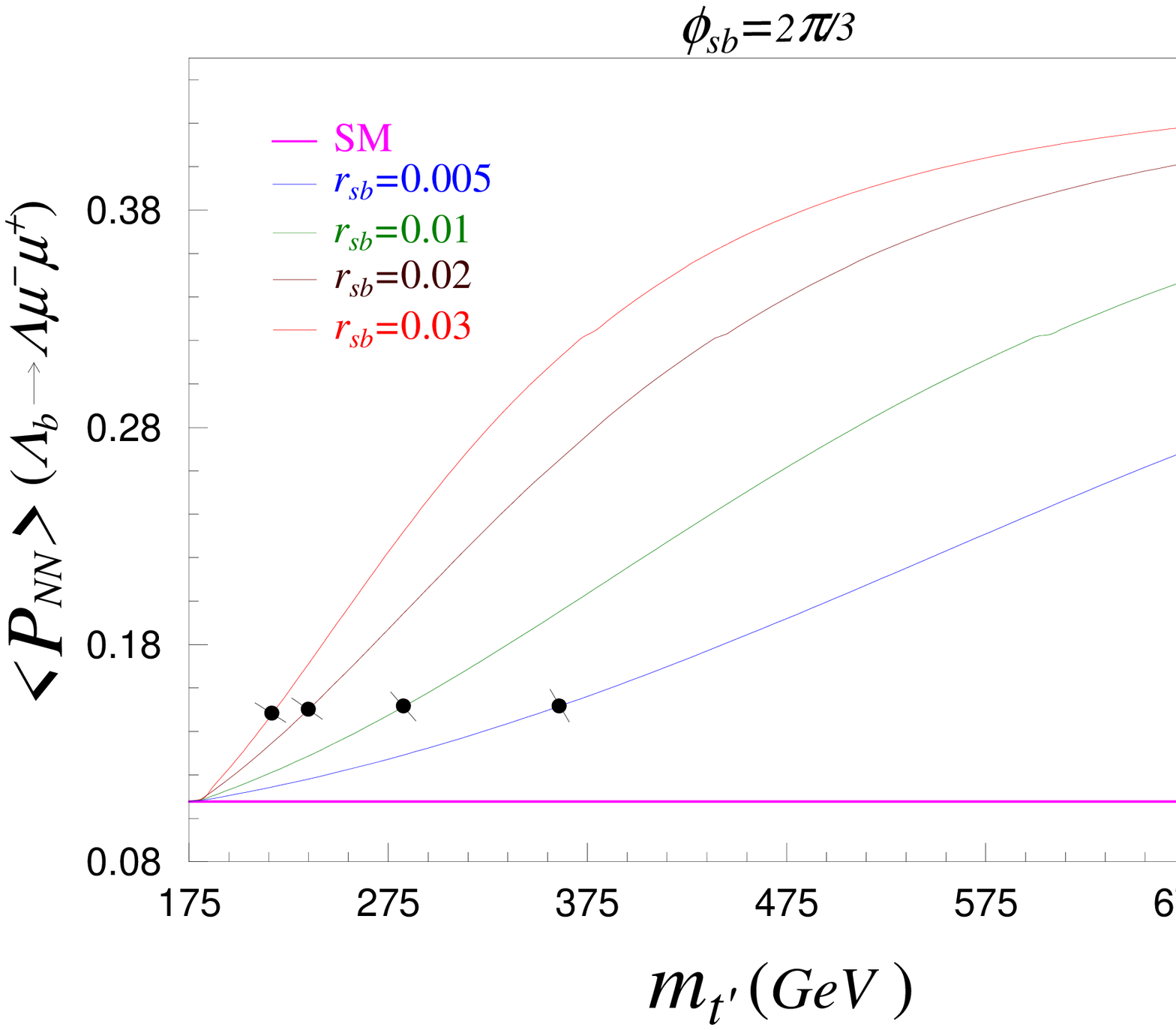}
\vskip 7.8cm \caption{}
\end{figure}

\begin{figure}
\vskip 2.5 cm
    \includegraphics{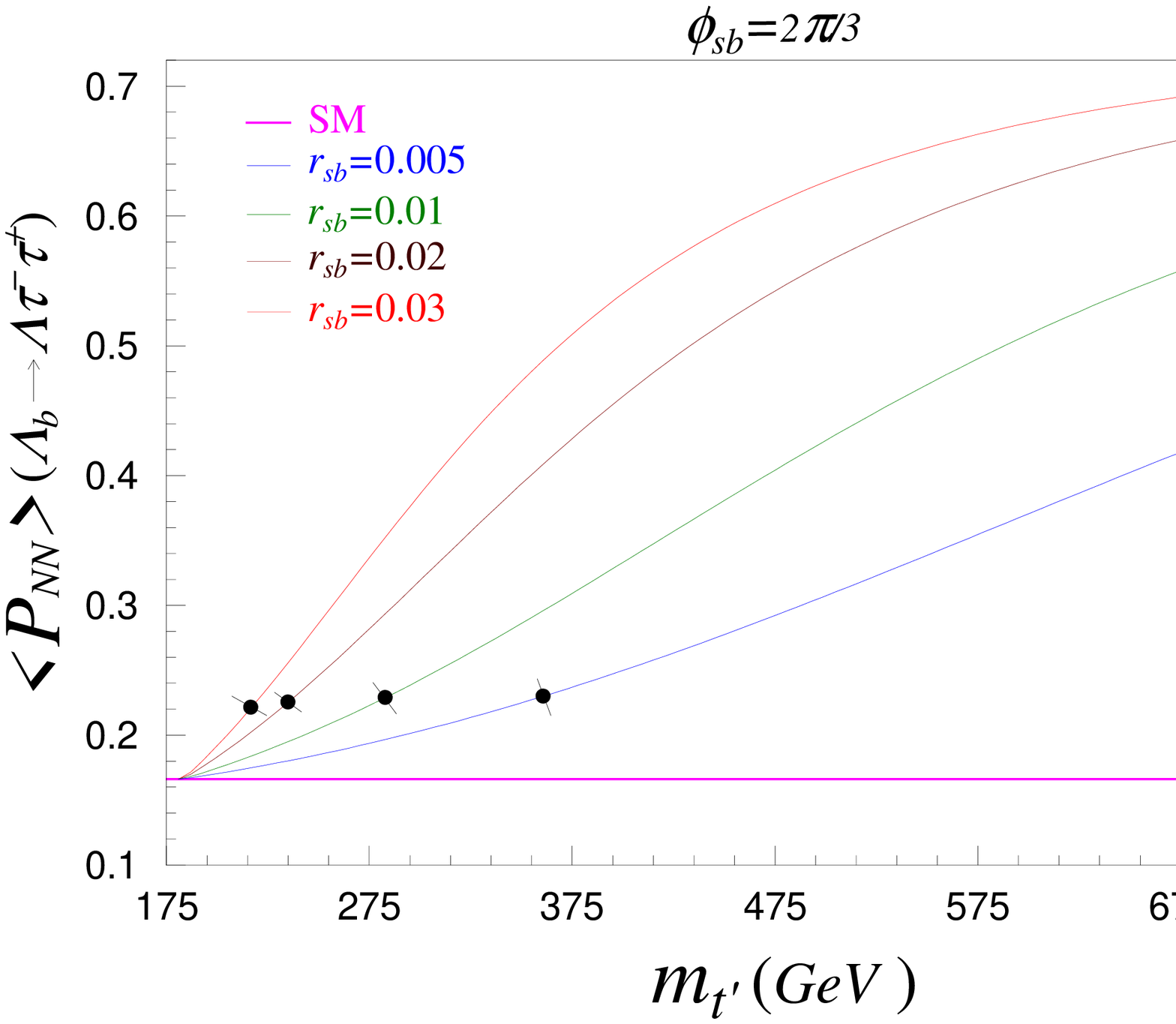}
\vskip 7.8cm \caption{}
\end{figure}

\end{document}